\documentclass[12pt]{article}
\RequirePackage[OT1]{fontenc}
\RequirePackage{amsthm,amsmath}
\usepackage[numbers,sort&compress]{natbib}

\usepackage{color, dsfont}
\usepackage{url}
\usepackage{amsmath,amsfonts,amsthm,amssymb,amsbsy,bbm}
\usepackage{multirow}
\usepackage{listings}
\usepackage[utf8]{inputenc}
\usepackage{amsmath}
\usepackage{amssymb}
\usepackage{bbm}
\usepackage{amsthm}
\usepackage{algorithm}
\usepackage{algorithmic}
\usepackage{graphicx}
\usepackage{array}
\usepackage{tablefootnote}
\usepackage{amssymb} 
\usepackage[dvipsnames]{xcolor}
\usepackage{array,booktabs,multirow,cleveref}
\usepackage{diagbox}

\newtheorem{theorem}{Theorem}[section]

\newtheorem{example}{Example}[section]

\begin{document}
	
\title{\large\bf An R Package AZIAD for Analyzing Zero-Inflated and Zero-Altered Data}
\author{Niloufar Dousti Mousavi$^{1}$, Hani Aldirawi$^{2}$ and Jie Yang$^{1}$\\
$^1$University of Illinois at Chicago and\\ $^2$California State University, San Bernardino}
	
\maketitle

\begin{abstract}
We introduce a newly developed R package AZIAD for analyzing zero-inflated or zero-altered data. Compared with existing R packages, AZIAD covers a much larger class of zero-inflated and hurdle models, including both discrete and continuous cases. It provides more accurate parameter estimates, along with the corresponding Fisher information matrix and confidence intervals. It achieves significantly larger power for model identification and selection. To facilitate the potential users, in this paper we provide detailed formulae and theoretical justifications for AZIAD, as well as new theoretical results on zero-inflated and zero-altered models. We use simulation studies to show the advantages of AZIAD functions over existing R packages and provide real data examples and executable R code to illustrate how to use our package for sparse data analysis and model selection.
\end{abstract}


\section{Introduction}\label{sec:introduction}

Sparse or zero-inflated data arise frequently from a rich variety of scientific disciplines including microbiome \citep{xia2018statistical, metwally2018review}, gene expression \citep{mcdavid2019graphical}, health care \citep{majo2011fixed}, insurance claim \citep{boucher2009number}, security \citep{chen2018bicycle}, and more. Modeling sparse data is very challenging due to the high proportion of zero values and the skewness of the distribution. Zero-inflated Poisson (ZIP), zero-inflated negative binomial (ZINB), Poisson hurdle (PH), and negative binomial hurdle (NBH) models have been widely used to model sparse data \citep{metwally2018review, aldirawi2019identifying, MetaLonDA}.

On one hand, more and more zero-inflated models have been proposed under different circumstances. On the other hand, it becomes more and more difficult for researchers and practitioners to choose the most appropriate model for their sparse data. Some partial comparisons among existing models have been done for certain sparse data. For examples, \cite{xu2015assessment} recommended ZINB and NBH models for microbiome data after comparing the performance of Poisson, ZIP, PH, NB (negative binomial), ZINB, and NBH models; while \cite{aldirawi2019identifying} indicated that two new models, zero-inflated beta negative binomial (ZIBNB) and beta negative binomial hurdle (BNBH) models, are more appropriate for microbiome data examples.

For analyzing zero-inflated data, there are some available {\tt R} packages from the Comprehensive R Archive Network (CRAN, \url{https://cran.r-project.org/}) including {\tt bzinb} \citep{bzinb}, {\tt hurdlr} \citep{hurdlr},  {\tt iZID} \citep{iZID}, {\tt gamlss} \citep{gamlss}, {\tt pscl} \citep{pscl}, 
{\tt mazeinda} \citep{mazeinda}, {\tt mhurdle} \citep{mhurdle}, {\tt rbtt} \citep{rbtt}, {\tt ZIBBSeqDiscovery} \citep{ZIBBSeqDiscovery}, {\tt ZIBseq} \citep{ZIBseq}, {\tt zic} \citep{zic}, {\tt ZIM} \citep{ZIM}, {\tt ziphsmm} \citep{ziphsmm}, etc. For example, {\tt iZID} covers 12 discrete distributions including Poisson, NB, BB (beta binomial), BNB (beta negative binomial) and their zero-inflated and hurdle versions. It implemented the bootstrapped Monte Carlo $p$-value estimates proposed by \cite{aldirawi2019identifying} for identifying a discrete distribution. For packages other than {\tt iZID}, please see \cite{wang2020identifying} for a good review.

The existing literature and packages on analyzing zero-inflated data, including the very recent work \cite{aldirawi2019identifying} and \cite{wang2020identifying}, are still limited for three reasons. First, only a small number of zero-inflated models were under consideration at the same time. Discrete and continuous baseline distributions were typically considered separately. Secondly, the accuracy of the parameter estimates and the power of the tests still have room for improvement. Thirdly, the confidence intervals of the parameter values were seldom provided along with their estimates, which is critical for model diagnostics, such as, testing whether the inflation or deflation of zeros exists in the data. 

In this paper, we introduce our newly developed {\tt R} package named {\tt AZIAD} for \underline{A}nalyzing \underline{Z}ero-\underline{I}nflated and Zero-\underline{A}ltered \underline{D}ata, available from the Comprehensive R Archive Network (CRAN, \url{https://cran.r-project.org/package=AZIAD}). Compared with the existing {\tt R} packages for similar purposes, our {\tt AZIAD} achieves the following significant improvements: (1) We not only cover discrete baseline distributions including Poisson, geometric, NB, BB, and BNB, but also cover commonly used continuous baseline distributions including normal (or Gaussian), log-normal, half-normal, and exponential distributions along with their zero-inflated and zero-altered (also known as hurdle) models; (2) By more precise specifications on solution forms under different situations and lower/upper bounds needed for numerical optimizations, our {\tt R} functions provide more accurate maximum likelihood estimates (MLE) even under extreme circumstances, which further gains more power when identifying the most appropriate zero-inflated or hurdle models for a given data set; (3) Following \cite{aldirawi2022modeling}, we provide not only MLEs for parameters, but also the Fisher information matrix and the corresponding confidence intervals for estimated parameters, which will facilitate the potential users from biological sciences, insurance, health studies, security, ecology, etc, to identify the most appropriate probabilistic model for their dataset and make robust statistical inference based on it.

The rest of this paper is organized as follows. In Section~\ref{sec:zamodel}, we not only review relevant theoretical results from \cite{aldirawi2022modeling}, but also provide new formulae for Fisher information matrices of the zero-inflated (ZI) and zero-altered (ZA or hurdle) models with geometric, BB, BNB, normal, log-normal, half-normal and exponential distributions, as well as ZINB. In Section~\ref{sec:numerical_study}, we summarize and use numerical examples to illustrate the improvements by using our package over existing {\tt R} packages, as well as identifying the most appropriate models for real data. We interpret and discuss indistinguishable pairs of distributions in Section~\ref{sec:discussion}.

\section{MLE and Fisher Information for ZI and ZA Models }\label{sec:zamodel}

\subsection{MLE and Fisher information for zero-altered or hurdle models}\label{sec:ZAP}
	
Zero-altered models or hurdle models have been widely used for modeling data with an excess or deficit of zeros (see, for example, \cite{metwally2018review}, for a good review). A general hurdle model consists a baseline distribution and a component generating the zeros. The baseline distribution could be fairly general with the distribution function $f_{\boldsymbol\theta}(y)$ and its zero-truncated version $f_{\rm tr}(y\mid \boldsymbol\theta) = [1 - p_0(\boldsymbol\theta)]^{-1} f_{\boldsymbol\theta}(y)$, $y\neq 0$, where ${\boldsymbol\theta}$ is the model parameter(s), and $p_0(\boldsymbol\theta) = P_{\boldsymbol\theta}(Y=0)$ is the probability that $Y=0$ under the baseline distribution. Following \cite{aldirawi2022modeling}, the distribution function of the  corresponding hurdle model can written as:
\begin{equation}\label{eq:hurdle}
	f_{\rm ZA}(y\mid \phi, \boldsymbol\theta) = \phi {\mathbf 1}_{\{y=0\}} + (1-\phi) f_{\rm tr}(y\mid \boldsymbol\theta) {\mathbf 1}_{\{y\neq 0\}} 
\end{equation}
where $\phi \in [0,1]$ is the weight parameter of zeros. Actually, $\phi=P(Y=0)$ if $Y\sim f_{\rm ZA}$.

If the baseline distribution is discrete with a probability mass function (pmf) $f_{\boldsymbol\theta}(y)$, such as Poisson, negative binomial (NB), geometric (Ge), beta binomial (BB) and beta negative binomial (BNB) distributions, then both $f_{\rm tr}(y\mid \boldsymbol\theta)$ and $f_{\rm ZA}(y\mid \phi, \boldsymbol\theta)$ are pmfs as well. The corresponding zero-altered or hurdle models may be called as zero-altered Poisson (ZAP) or Poisson hurdle (PH), zero-altered negative binomial (ZANB) or negative binomial hurdle (NBH), zero-altered geometric (ZAGe) or geometric hurdle (GeH), zero-altered beta binomial (ZABB) or beta binomial hurdle (BBH), zero-altered beta negative binomial (ZABNB) or beta negative binomial hurdle (BNBH) models, respectively.
	
If the baseline distribution is continuous with a probability density function (pdf) $f_{\boldsymbol\theta}(y)$, such as	Gaussian (or normal), log-normal, half-normal and exponential distributions, then $p_0(\boldsymbol\theta)=0$ and $f_{\rm tr}(y\mid \boldsymbol\theta) = f_{\boldsymbol\theta}(y)$. In this case, $f_{\rm ZA}(y\mid \phi, \boldsymbol\theta)$ in \eqref{eq:hurdle} represents a mixture distribution consisting of a discrete component at $0$ with probability $\phi$ and a continuous component with density function $(1-\phi) f_{\boldsymbol\theta}(y)$. We will revisit these continuous cases in Section~\ref{sec:za_zi}. In this section, we focus on discrete cases.

The parameters of the hurdle model~\eqref{eq:hurdle} include both $\phi$ and $\boldsymbol\theta$. In this paper, we adopt the maximum likelihood estimate (MLE) for estimating the parameters (see, for example, Section~1.3.1 in \cite{agresti2013}, for justifications on adopting MLE). Let $Y_1, \ldots, Y_n$ be a random sample from model~\eqref{eq:hurdle}. Then the likelihood function of $(\phi, \boldsymbol\theta)$ is
\begin{equation}\label{eq:hurdelmle}
	L(\phi, \boldsymbol\theta) = \phi^{n-m} (1-\phi)^m \cdot  \prod_{i:Y_i\neq 0} f_{\rm tr}(Y_i\mid \boldsymbol\theta)
\end{equation}
where $m=\#\{i:Y_i\neq 0\}$ is the number of nonzero observations. According to Theorem~1 in \cite{aldirawi2022modeling}, the maximum likelihood estimate (MLE) of $(\phi, \boldsymbol\theta)$ that maximizes \eqref{eq:hurdelmle} is
\begin{equation}\label{eq:hurdlemle}
\hat{\phi}=1-\frac{m}{n},\>\>\> 
		\hat{\boldsymbol\theta}={\rm argmax}_{\boldsymbol\theta} \prod_{i:Y_i\neq 0} f_{\rm tr}(Y_i\mid \boldsymbol\theta)
\end{equation}
That is, $\hat{\boldsymbol\theta}$ is simply the MLE for the truncated model with distribution function $f_{\rm tr}(y\mid \boldsymbol\theta)  =  f_{\boldsymbol\theta}(y)/[1-p_0(\boldsymbol\theta)], y\neq 0$.

According to Theorem~3 in \cite{aldirawi2022modeling},
under some regularity conditions, the Fisher information matrix of the zero-altered distribution is
\begin{equation}\label{eq:hurdle_fisher}
{\mathbf F}_{\rm ZA} = 
\left[\begin{array}{cc}
\phi^{-1} (1-\phi)^{-1} & {\mathbf 0}^T\\
{\mathbf 0} & {\mathbf F}_{\rm ZA\boldsymbol\theta}
\end{array}\right]
\end{equation}
where 
\[
{\mathbf F}_{\rm ZA\boldsymbol\theta} = -\frac{1-\phi}{1-p_0(\boldsymbol\theta)} \left(E\left[\frac{\partial^2 \log f_{\boldsymbol\theta}(Y')}{\partial \boldsymbol\theta \partial \boldsymbol\theta^T} \right] + \frac{p_0(\boldsymbol\theta)}{1-p_0(\boldsymbol\theta)} \cdot \frac{\partial \log p_0(\boldsymbol\theta)}{\partial \boldsymbol\theta}\cdot \frac{\partial \log p_0(\boldsymbol\theta)}{\partial \boldsymbol\theta^T}\right)
\]
and $Y'$ follows the baseline distribution $f_{\boldsymbol\theta}(y)$. Note that the sample size $n$ does not show up in \eqref{eq:hurdle_fisher} since the Fisher information here is of the distribution, not of the sample.

Major advantages for adopting MLE \eqref{eq:hurdlemle} and calculating the Fisher information matrix \eqref{eq:hurdle_fisher} include: (i) $\hat\phi$ and $\hat{\boldsymbol\theta}$ are consistent estimators of $\phi$ and $\boldsymbol\theta$, respectively (Theorem~2 in \cite{aldirawi2022modeling}); (ii) $\sqrt{n}(\hat\phi - \phi)  \stackrel{{\cal L}}{\rightarrow} N(0, \phi (1-\phi))$ and $\sqrt{n}(\hat{\boldsymbol\theta} - \boldsymbol\theta)  \stackrel{{\cal L}}{\rightarrow} N\left({\mathbf 0}, {\mathbf F}_{\rm ZA\boldsymbol\theta}^{-1} \right)$, which provides the formulae for building up approximate confidence intervals and relevant hypothesis tests for $\phi$ and $\boldsymbol\theta$ (see, for example, Sections~1.3.3 and 1.3.4 in \cite{agresti2013}). For example, an approximate $(1-\alpha)100\%$ confidence interval for $\phi$ is
\begin{equation}\label{eq:hurdle_phi_ci}
\phi \in \left(\hat\phi - \frac{z_{\frac{\alpha}{2}}}{\sqrt{n}}  \sqrt{\hat\phi (1-\hat\phi)}, \ \hat\phi + \frac{z_{\frac{\alpha}{2}}}{\sqrt{n}}  \sqrt{\hat\phi (1-\hat\phi)}\right)
\end{equation}
where $z_{\frac{\alpha}{2}} = \Phi^{-1}(1-\frac{\alpha}{2})$ is the upper $\frac{\alpha}{2}$th quantile of the standard normal distribution, and $\alpha \in (0,1)$ is the desired significance level, such as $0.05$. If $p_0(\boldsymbol\theta)$ does not belong to the confidence interval~\eqref{eq:hurdle_phi_ci}, then there is a significant evidence for zero-inflation or deflation.

Therefore, calculating MLE and Fisher information accurately and efficiently is the basis of sound and thorough statistical inference. 

Explicit formulae of the gradients ($\partial\log f_{\boldsymbol\theta}(y)/\partial \boldsymbol\theta$ and $\partial\log p_0(\boldsymbol\theta)/\partial\boldsymbol\theta$, need for finding MLEs) and the Fisher information matrices of zero-altered Poisson (ZAP) or Poisson hurdle (PH) model, zero-altered negative binomial (ZANB) or negative binomial hurdle (NBH) model, have been provided in Examples~2 and 3 of \cite{aldirawi2022modeling}, respectively. In this section, we will provide explicit formulae for zero-altered geometric, beta binomial, and beta negative binomial models. 

\begin{example}\label{ex:ZAGEOM}{\bf Zero-altered geometric (ZAGe) or geometric hurdle (GeH) model} 
{\rm has been used, for example, in \cite{mullahy1986specification} for modeling count data in econometrics. The pmf of the baseline distribution can be written as $f_{p} (y) = p(1-p)^{y}$ with $y \in \{0, 1, 2, \ldots\}$ and parameter $p\in (0,1)$. In this case, $p_0(p) = p$, $\frac{\partial\log f_{p}(y)}{\partial p} = \frac{1}{p} - \frac{y}{1-p}$, and $\frac{\partial\log p_{0}(p)}{\partial p} = \frac{1}{p}$.

According to Theorem~3 in \cite{aldirawi2022modeling}, the Fisher information matrix of the ZAGe or GeH distribution can be written as
\[
{\mathbf F}_{\rm ZAGe}=\begin{bmatrix}
\frac{1}{\phi(1-\phi)} & 0 \\
0 & \frac{1-\phi}{p^2(1-p)}
\end{bmatrix}
\]
}
\hfill{$\Box$}
\end{example}

\begin{example}\label{ex:ZABB}{\bf Zero-altered beta-binomial (ZABB) or beta-binomial hurdle (BBH) model}
{\rm
has been used by \cite{aldirawi2019identifying} for modeling microbiome data. The pmf of the baseline distribution with parameters  $\boldsymbol\theta = (n, \alpha, \beta)\in \mathbb{N}\times (0,\infty) \times (0,\infty)$  is given by $f_{\boldsymbol\theta} (y) = {n\choose y} \frac{{\rm Beta}(y + \alpha, n - y + \beta)}{{\rm Beta}(\alpha,\beta)}$, $y \in \{0, 1, 2, \ldots, n\}$. In this case, $p_0(\boldsymbol\theta) = \frac{\Gamma(n + \beta) \Gamma(\alpha + \beta)}{\Gamma(n + \alpha + \beta)  \Gamma(\beta)}$. Then
\begin{eqnarray*}
	\frac{\partial\log f_{\boldsymbol\theta}(y)}{\partial n} &=& \Psi(n+1) - \Psi(n-y+1) + \Psi(n-y+\beta) - \Psi(n+\alpha+\beta)\\
	\frac{\partial\log f_{\boldsymbol\theta}(y)}{\partial \alpha} &=& \Psi(y+\alpha) - \Psi(n+\alpha+\beta) + \Psi(\alpha+\beta) - \Psi(\alpha)\\
	\frac{\partial\log f_{\boldsymbol\theta}(y)}{\partial \beta} &=& \Psi(n-y+\beta) - \Psi(n+\alpha+\beta) +  \Psi(\alpha+\beta) - \Psi(\beta) \\
	\frac{\partial \log p_0(\boldsymbol\theta)}{\partial n} &=& \Psi(n+\beta) - \Psi(n+\alpha+\beta) \\
	\frac{\partial \log p_0(\boldsymbol\theta)}{\partial \alpha} &=& \Psi(\alpha+\beta) - \Psi(n+\alpha+\beta) \\
	\frac{\partial\log p_0(\boldsymbol\theta)}{\partial \beta} &=& \Psi(n+\beta) + \Psi(\alpha+\beta) - \Psi(n+\alpha+\beta) - \Psi(\beta)
\end{eqnarray*}
where $\Psi(\cdot) = \Gamma'(\cdot)/\Gamma(\cdot)$ is known as the {\it digamma} function. Note that the range of parameter $n$ can be extended to positive real numbers. According to Theorem~3 in \cite{aldirawi2022modeling} , the Fisher information matrix of the ZABB or BBH distribution is
\begin{eqnarray*}
{\mathbf F}_{\rm ZABB}=\begin{bmatrix}
\frac{1}{\phi(1-\phi)} & {\mathbf 0}^T \\
{\mathbf 0} & {\mathbf F}_{{\rm BB}\boldsymbol\theta}
\end{bmatrix}
\end{eqnarray*}
where 
\begin{eqnarray*}
{\mathbf F}_{{\rm BB}\boldsymbol\theta} &=& -\frac{(1-\phi)\Gamma(n+\alpha+\beta)\Gamma(\beta)}{\Gamma(n+\alpha+\beta)\Gamma(\beta)-\Gamma(n+\beta)\Gamma(\alpha+\beta)} \left(\begin{bmatrix}
A_{11} & A_{12} & A_{13}\\
A_{12} & A_{22} & A_{23}\\
A_{13} & A_{23} & A_{33}
\end{bmatrix}\right.\\
& & +\
\left.\frac{\Gamma(n + \beta) \Gamma(\alpha + \beta)}{\Gamma(n + \alpha + \beta) \Gamma(\beta) - \Gamma(n + \beta) \Gamma(\alpha + \beta)}
\begin{bmatrix}
B_{11} & B_{12} & B_{13}\\
B_{12} & B_{22} & B_{23}\\
B_{13} & B_{23} & B_{33}
\end{bmatrix}\right)
\end{eqnarray*}
with 
\begin{eqnarray*}
A_{11} &=& \Psi_{1}(n+1)-\Psi_{1}(n+\alpha+\beta)+E\Psi_{1}(n-Y'+\beta) - E\Psi_{1}(n-Y'+1)\\
A_{12} &=& - \Psi_{1}(n+\alpha+\beta)\\
A_{13} &=& E\Psi_{1}(n-Y'+\beta)-\Psi_{1}(n+\alpha+\beta)\\
A_{22} &=& \Psi_{1}(\alpha+\beta)-\Psi_{1}(n+\alpha+\beta)-\Psi_{1}(\alpha)+E\Psi_{1}(Y'+\alpha)\\
A_{23} &=& \Psi_{1}(\alpha+\beta)-\Psi_{1}(n+\alpha+\beta)\\
A_{33} &=& \Psi_{1}(\alpha+\beta)-\Psi_{1}(\beta)-\Psi_{1}(n+\alpha+\beta)+E\Psi_{1}(n-Y'+\beta)\\
B_{11}&=&[\Psi(n+\beta) - \Psi(n+\alpha+\beta)]^2\\
B_{12}&=&[\Psi(n+\beta) - \Psi(n+\alpha+\beta)]
\cdot[\Psi(\alpha+\beta) - \Psi(n+\alpha+\beta)]\\
B_{13}&=&[\Psi(n+\beta) - \Psi(n+\alpha+\beta)]\cdot[\Psi(n+\beta) + \Psi(\alpha+\beta) - \Psi(n+\alpha+\beta) - \Psi(\beta)]\\
B_{22}&=&[\Psi(\alpha+\beta)-\Psi(n+\alpha+\beta)]^2\\
B_{23}&=&[\Psi(\alpha+\beta) - \Psi(n+\alpha+\beta)]\cdot[\Psi(n+\beta) + \Psi(\alpha+\beta) - \Psi(n+\alpha+\beta) - \Psi(\beta)]\\
B_{33}&=&[\Psi(n+\beta) + \Psi(\alpha+\beta) - \Psi(n+\alpha+\beta) - \Psi(\beta)]^2
\end{eqnarray*}
where $\Psi_1(\cdot) = \Psi'(\cdot)$ is known as the {\it trigamma} function, and $Y'$ follows the baseline distribution $f_{\boldsymbol\theta}(y)$.  
}
\hfill{$\Box$}
\end{example}

Due to the lack of simple forms, we may use Monte Carlo estimates for calculating $E\Psi_1(\cdot)$ in Example~\ref{ex:ZABB} and others approximately. For example, $E\Psi_1(n-Y'+\beta)$ can be approximated by $N^{-1} \sum_{i=1}^N \Psi_1(n-Y_i'+\beta)$ with simulated $Y_1', \ldots, Y_N'$ from $f_{\boldsymbol\theta}(y)$.

Another computational issue involved in Example~\ref{ex:ZABB} and others is that $\log[\Gamma(n+\alpha+\beta) \Gamma(\beta) - \Gamma(n+\beta) \Gamma(\alpha+\beta)]$, which is relevant to $\log p_0(\boldsymbol\theta)$, may be undefined numerically for large $n$ since both $\Gamma(n+\alpha+\beta)$ and $\Gamma(n+\beta)$ are numerical infinity. To overcome this kind of issues, we use the fact $\log(A-B)=\log(1 - \exp(\log B-\log A)) + \log A$ if $A \geq B$. It can be verified that $\Gamma(n+\alpha+\beta) \Gamma(\beta) > \Gamma(n+\beta) \Gamma(\alpha+\beta)$.

\begin{example}\label{ex:ZABNB}{\bf Zero-altered beta negative binomial (ZABNB) or beta negative binomial hurdle (BNBH) model} {\rm has been recommended by \cite{aldirawi2019identifying} and \cite{aldirawi2022modeling} for modeling microbiome data. The pmf of the baseline distribution with parameters $\boldsymbol\theta = (r, \alpha, \beta) \in \mathbb{N} \times (0,\infty) \times (0,\infty)$ is given by $f_{\boldsymbol\theta}(y) = {r+y-1\choose y} \frac{{\rm Beta}(r + \alpha, y + \beta)}{{\rm Beta}(\alpha, \beta)}$, $y \in \{0, 1, 2, \ldots\}$. Then $p_0(\boldsymbol\theta) =\frac{\Gamma(r + \alpha) \Gamma(\alpha+\beta)}{\Gamma(r + \alpha + \beta) \Gamma(\alpha)}$. Note that $r$ can also be extended to positive real numbers.  
\begin{eqnarray*}
	\frac{\partial\log f_{\boldsymbol\theta}(y)}{\partial r} &=& \Psi(r + y) - \Psi(r) + \Psi(r + \alpha) - \Psi(r+y+\alpha+\beta)\\
	\frac{\partial\log f_{\boldsymbol\theta}(y)}{\partial \alpha} &=& \Psi(r+\alpha) - \Psi(r+y+\alpha+\beta) + \Psi(\alpha+\beta) - \Psi(\alpha)\\
	\frac{\partial\log f_{\boldsymbol\theta}(y)}{\partial \beta} &=& \Psi(y+\beta) - \Psi(r+y+\alpha+\beta) +  \Psi(\alpha+\beta) - \Psi(\beta) \\
	\frac{\partial \log p_0(\boldsymbol\theta)}{\partial r} &=& \Psi(r+\alpha) - \Psi(r+\alpha+\beta) \\
	\frac{\partial \log p_0(\boldsymbol\theta)}{\partial \alpha} &=& \Psi(r+\alpha) + \Psi(\alpha+\beta) - \Psi(r+\alpha+\beta) - \Psi(\alpha) \\
	\frac{\partial\log p_0(\boldsymbol\theta)}{\partial \beta} &=& \Psi(\alpha+\beta) - \Psi(r+\alpha+\beta)
\end{eqnarray*}
According to Theorem~3 in \cite{aldirawi2022modeling}, the Fisher information matrix of the ZABNB or BNBH distribution is
\begin{eqnarray*}
{\mathbf F}_{\rm ZABNB}=\begin{bmatrix}
\frac{1}{\phi(1-\phi)} & {\mathbf 0}^T \\
{\mathbf 0} & {\mathbf F}_{{\rm BNB}\boldsymbol\theta}
\end{bmatrix}\\
\end{eqnarray*}
where 
\begin{eqnarray*}
{\mathbf F}_{{\rm BNB}\boldsymbol\theta} &=& -\frac{(1-\phi)\Gamma(r+\alpha+\beta)\Gamma(\alpha)}{\Gamma(r+\alpha+\beta)\Gamma(\alpha)-\Gamma(r+\alpha)\Gamma(\alpha+\beta)} \left(\begin{bmatrix}
A_{11} & A_{12} & A_{13}\\
A_{12} & A_{22} & A_{23}\\
A_{13} & A_{23} & A_{33}
\end{bmatrix}\right.\\
& & +\
\left.\frac{\Gamma(r+\alpha)\Gamma(\alpha+\beta)}{\Gamma(r+\alpha+\beta)\Gamma(\alpha)-\Gamma(r+\alpha)\Gamma(\alpha+\beta)}
\begin{bmatrix}
B_{11} & B_{12} & B_{13}\\
B_{12} & B_{22} & B_{23}\\
B_{13} & B_{23} & B_{33}
\end{bmatrix}\right)
\end{eqnarray*}
with
\begin{eqnarray*}
A_{11}&=&E\Psi_{1}(r + Y') - \Psi_{1}(r) + \Psi_{1}(r + \alpha) - E\Psi_{1}(r+Y'+\alpha+\beta)\\
A_{12}&=&\Psi_{1}(r+\alpha)-E\Psi_{1}(r+Y'+\alpha+\beta)\\
A_{13}&=&-E\Psi_{1}(r+Y'+\alpha+\beta)\\
A_{22}&=&\Psi_{1}(r+\alpha) - E\Psi_{1}(r+Y'+\alpha+\beta) + \Psi_{1}(\alpha+\beta) - \Psi_{1}(\alpha)\\
A_{23}&=&-E\Psi_{1}(r+Y'+\alpha+\beta)+\Psi_{1}(\alpha+\beta)\\
A_{33}&=&E\Psi_{1}(Y'+\beta) - E\Psi_{1}(r+Y'+\alpha+\beta) +  \Psi_{1}(\alpha+\beta) - \Psi_{1}(\beta) \\
B_{11}&=&[\Psi(r+\alpha)-\Psi(r+\alpha+\beta)]^2\\
B_{12}&=&[\Psi(r+\alpha)-\Psi(r+\alpha+\beta)]\cdot[\Psi(r+\alpha)+\Psi(\alpha+\beta)-\Psi(r+\alpha+\beta)-\Psi(\alpha)]\\
B_{13}&=&[\Psi(r+\alpha)-\Psi(r+\alpha+\beta)]\cdot[\Psi(\alpha + \beta)-\Psi(r+\alpha+\beta)]\\
B_{22}&=&[\Psi(r+\alpha)+\Psi(\alpha+\beta)-\Psi(r+\alpha+\beta)-\Psi(\alpha)]^2\\
B_{23}&=&[\Psi(r+\alpha)+\Psi(\alpha+\beta)-\Psi(r+\alpha+\beta)-\Psi(\alpha)]\cdot[\Psi(\alpha + \beta)-\Psi(r+\alpha+\beta)]\\
B_{33}&=&[\Psi(\alpha + \beta)-\Psi(r+\alpha+\beta)]^2
\end{eqnarray*}
As for $\Psi_1$ and relevant calculations, please see the arguments right after Example~\ref{ex:ZABB}.
}
\hfill{$\Box$}
\end{example}
	
\subsection{MLE and Fisher information for zero-inflated models}\label{sec:zimle}
	
When data is sparse, a zero-inflated (ZI) model is more commonly used in practice, which assumes an excess of zeros (see, for example, \cite{metwally2018review} for a good review). Similar as the zero-altered (ZA) models in Section~\ref{sec:ZAP}, there is a baseline distribution with distribution function $f_{\boldsymbol\theta}(y)$ and parameter(s) $\boldsymbol\theta$. We also denote $p_0(\boldsymbol\theta) = P_{\boldsymbol\theta}(Y=0)$, if $Y\sim f_{\boldsymbol\theta}$. Different from ZA models, a zero-weighting parameter $\phi \in [0,1]$ adds additional probability of zeros to the ZI model. Following \cite{aldirawi2022modeling}, we write the distribution function of the corresponding ZI model as
\begin{equation}\label{eq:zimodel}
	f_{\rm ZI}(y\mid \phi, {\boldsymbol\theta}) = [\phi + (1-\phi)p_0(\boldsymbol\theta)] {\mathbf 1}_{\{y=0\}} + (1-\phi) f_{{\boldsymbol\theta}}(y) {\mathbf 1}_{\{y\neq 0\}}
	\end{equation}
Note that if $Y\sim f_{\rm ZI}$, then $P(Y=0) = \phi + (1-\phi)p_0(\boldsymbol\theta)$, which is larger than $\phi$ in general. 

When the baseline distribution is either continuous with a pdf $f_{\boldsymbol\theta}(y)$ or discrete but with $p_0(\boldsymbol\theta)=0$, the corresponding zero-inflated model \eqref{eq:zimodel} is essentially the same as the corresponding zero-altered model \eqref{eq:hurdle}. Examples include Gaussian or normal, log-normal, half-normal, and exponential distributions. We will revisit them in Section~\ref{sec:za_zi}.

When the baseline distribution is discrete, such as Poisson (P), negative binomial (NB), geometric (Ge), beta binomial (BB) and beta negative binomial (BNB), the corresponding zero-inflated models can be written as ZIP, ZINB, ZIGe, ZIBB, and ZIBNB, respectively.

Given a random sample $Y_1, \ldots, Y_n$ from the zero-inflated model $f_{\rm ZI}(y|\phi, \boldsymbol\theta)$, the likelihood function of $(\phi, \boldsymbol\theta)$ can be written as
\begin{eqnarray}
	L(\phi, \boldsymbol{\theta}) 
		&=& \left[\phi+p_0({\boldsymbol\theta}) (1-\phi)\right]^{n-m}\cdot  (1-\phi)^{m}
		\prod_{i: Y_i \neq 0}f_{\boldsymbol\theta}(Y_i)\nonumber\\
		&=& \left[\phi+p_0({\boldsymbol\theta}) (1-\phi)\right]^{n-m}\cdot  (1-\phi)^{m}
		\left(1 - p_0(\boldsymbol\theta)\right)^{m} \cdot \prod_{i: Y_i \neq 0}f_{\rm tr}(Y_{i}\mid \boldsymbol\theta)\label{eq:zi_likelihood}
\end{eqnarray}
where $m= \#\{i:Y_i \neq 0\}$, and $f_{\rm tr}(y; \boldsymbol\theta) = f_{\boldsymbol\theta}(y)/[1-p_0(\boldsymbol\theta)], y\neq 0$.

According to Theorem~4 in \cite{aldirawi2022modeling}, the maximum likelihood estimate $(\hat\phi, \hat{\boldsymbol\theta})$ maximizing \eqref{eq:zi_likelihood} can be obtained as follows: 
\begin{itemize}
    \item[(0)] Determine $\boldsymbol\theta_* = {\rm argmax}_{\boldsymbol\theta} L_{\rm tr}(\boldsymbol\theta)$, where $L_{\rm tr}(\boldsymbol\theta) = \prod_{i: Y_i \neq 0} f_{\rm tr}(Y_i; \boldsymbol\theta)$. 
	\item[(1)] If $m/n \leq 1-p_0({\boldsymbol\theta_*})$, then $\hat{\boldsymbol\theta} = \boldsymbol\theta_*$ and $\hat\phi = 1 -  [1-p_0({\boldsymbol\theta_*})]^{-1} \cdot m/n$.
	\item[(2)] Otherwise, $\hat{\boldsymbol\theta} = {\rm argmax}_{\boldsymbol\theta} L(\psi(\boldsymbol\theta), \boldsymbol\theta)$ and $\hat{\phi} = 1 - \psi(\hat{\boldsymbol\theta}) \cdot [1-p_0(\hat{\boldsymbol\theta})]^{-1}$, where $\psi(\boldsymbol\theta) = \min\{m/n, 1-p_0({\boldsymbol\theta})\}$, and $L(\psi, \boldsymbol\theta) =  (1-\psi)^{n-m} \psi^m  \prod_{i: Y_i \neq 0} f_{\rm tr}(Y_i; \boldsymbol\theta)$.
\end{itemize}
Solving for the MLE $(\hat\phi, \hat{\boldsymbol\theta})$ may involve two maximization problems, $\boldsymbol\theta_* = {\rm argmax}_{\boldsymbol\theta} L_{\rm tr}(\boldsymbol\theta)$ and $\hat{\boldsymbol\theta} = {\rm argmax}_{\boldsymbol\theta} L(\psi(\boldsymbol\theta), \boldsymbol\theta)$. The following first order derivatives may be needed by, for example, quasi-Newton algorithms:
\begin{eqnarray*}
		\frac{\partial \log L_{\rm tr} ( \boldsymbol\theta)}{\partial \boldsymbol\theta} &=&  \sum_{i:Y_i \neq 0} \frac{\partial\log f_{\boldsymbol\theta}(Y_i)}{\partial \boldsymbol\theta} - m \frac{\partial\log[1-p_0(\boldsymbol\theta)]}{\partial \boldsymbol\theta}\\ 
		\frac{\partial \log L(\psi(\boldsymbol\theta), \boldsymbol\theta)}{\partial \boldsymbol\theta}  &=& \left\{\begin{array}{cl} 
		\sum_{i:Y_i\neq 0} \frac{\partial\log f_{\boldsymbol\theta}(Y_i)}{\partial \boldsymbol\theta} - m\frac{\partial\log[1-p_0(\boldsymbol\theta)]}{\partial \boldsymbol\theta} & \mbox{ if }1-p_0(\boldsymbol\theta) > \frac{m}{n}\\
        \sum_{i:Y_i\neq 0} \frac{\partial\log f_{\boldsymbol\theta}(Y_i)}{\partial \boldsymbol\theta} + (n-m) \frac{\partial\log p_0(\boldsymbol\theta)}{\partial \boldsymbol\theta} & \mbox{ if } 1-p_0(\boldsymbol\theta) < \frac{m}{n}
        \end{array}\right.
\end{eqnarray*}
Note that
\[
	\frac{\partial\log[1-p_0(\boldsymbol\theta)]}{\partial \boldsymbol\theta} = - \frac{p_0(\boldsymbol\theta)}{1-p_0(\boldsymbol\theta)}\cdot \frac{\partial\log p_0(\boldsymbol\theta)}{\partial \boldsymbol\theta}
\]
Thus for different models, we only need to prepare specific formulae of $\partial\log f_{\boldsymbol\theta}(y)/\partial \boldsymbol\theta$ and $\partial \log p_0(\boldsymbol\theta)/\partial\boldsymbol\theta$ for numerical calculations.

According to Theorem~5 in \cite{aldirawi2022modeling}, under some regularity conditions, the Fisher information matrix of a general zero-inflated distribution is
\begin{equation}\label{eq:fisher_ZI}
{\mathbf F}_{\rm ZI} = 
\left[\begin{array}{cc}
\frac{1-p_0(\boldsymbol\theta)}{[\phi + (1-\phi)p_0(\boldsymbol\theta)] (1-\phi)} & \frac{p_0(\boldsymbol\theta)}{\phi + (1-\phi) p_0(\boldsymbol\theta)} \cdot \frac{\partial \log p_0(\boldsymbol\theta)}{\partial \boldsymbol\theta^T} \\
\frac{p_0(\boldsymbol\theta)}{\phi + (1-\phi) p_0(\boldsymbol\theta)} \cdot \frac{\partial \log p_0(\boldsymbol\theta)}{\partial \boldsymbol\theta}  & {\mathbf F}_{\rm ZI\boldsymbol\theta} 
\end{array}\right]
\end{equation}
where 
\[
{\mathbf F}_{\rm ZI\boldsymbol\theta} = -(1-\phi) \left(E\left[\frac{\partial^2 \log f_{\boldsymbol\theta}(Y')}{\partial \boldsymbol\theta \partial \boldsymbol\theta^T} \right] + \frac{\phi p_0(\boldsymbol\theta)}{\phi + (1-\phi) p_0(\boldsymbol\theta)} \cdot \frac{\partial \log p_0(\boldsymbol\theta)}{\partial \boldsymbol\theta}\cdot \frac{\partial \log p_0(\boldsymbol\theta)}{\partial \boldsymbol\theta^T}\right)
\]
and $Y'$ follows the baseline distribution $f_{\boldsymbol\theta}(y)$. We denote ${\mathbf F}_{\boldsymbol\theta} = -E\left[\frac{\partial^2 \log f_{\boldsymbol\theta}(Y')}{\partial \boldsymbol\theta \partial \boldsymbol\theta^T} \right]$, which is essentially the Fisher information matrix of the baseline distribution.

The Fisher information matrix \eqref{eq:fisher_ZI} is more complicated than \eqref{eq:hurdle_fisher} for zero-altered models. To obtain the asymptotic distributions of MLEs, it is more convenient to obtain the following theorem by applying the formulae (see, for example, Formulae~4.33 and 14.13(b) in \cite{seber2008}) of block matrices and the Sherman-Morrison formula (see, for example, Section~2.1.4 in \cite{golub2013}). 

\begin{theorem}\label{thm:inverse_fisher_ZI}
Suppose $\phi \in (0,1)$, $|{\mathbf F}_{\boldsymbol\theta}| \neq 0$ and $|{\mathbf F}_{\rm ZI}| \neq 0$. Then
\[
{\mathbf F}_{\rm ZI}^{-1} = \left[
\begin{array}{cc}
\phi (1-\phi)\cdot \frac{d_{\boldsymbol\theta} \delta_{\boldsymbol\theta}}{d_{\boldsymbol\theta}\delta_{\boldsymbol\theta} - p_0(\boldsymbol\theta)} & -\frac{\phi p_0(\boldsymbol\theta)}{d_{\boldsymbol\theta}\delta_{\boldsymbol\theta} - p_0(\boldsymbol\theta)}\frac{\partial\log p_0(\boldsymbol\theta)}{\partial \boldsymbol\theta^T} {\mathbf F}_{\boldsymbol\theta}^{-1}\\
-\frac{\phi p_0(\boldsymbol\theta)}{d_{\boldsymbol\theta}\delta_{\boldsymbol\theta} - p_0(\boldsymbol\theta)}{\mathbf F}_{\boldsymbol\theta}^{-1}\frac{\partial\log p_0(\boldsymbol\theta)}{\partial \boldsymbol\theta} & {\mathbf D}_{\boldsymbol\theta}\\
\end{array}\right]
\]
where $d_{\boldsymbol\theta} = \phi + (1-\phi) p_0(\boldsymbol\theta) > 0$, $\delta_{\boldsymbol\theta} = 1-\frac{\phi p_0(\boldsymbol\theta)}{d_{\boldsymbol\theta}} \frac{\partial\log p_0(\boldsymbol\theta)}{\partial \boldsymbol\theta^T} {\mathbf F}_{\boldsymbol\theta}^{-1} \frac{\partial\log p_0(\boldsymbol\theta)}{\partial \boldsymbol\theta} \neq 0$, and
\[
{\mathbf D}_{\boldsymbol\theta} = \frac{1}{1-\phi}\left[ {\mathbf F}_{\boldsymbol\theta}^{-1} + \frac{\phi p_0(\boldsymbol\theta)}{d_{\boldsymbol\theta} \delta_{\boldsymbol\theta} - p_0(\boldsymbol\theta)} {\mathbf F}_{\boldsymbol\theta}^{-1} \frac{\partial\log p_0(\boldsymbol\theta)}{\partial \boldsymbol\theta} \frac{\partial\log p_0(\boldsymbol\theta)}{\partial \boldsymbol\theta^T} {\mathbf F}_{\boldsymbol\theta}^{-1}\right]
\]
\hfill{$\Box$}
\end{theorem}

The proof of Theorem~\ref{thm:inverse_fisher_ZI} is relegated to the Supplementary Materials (Section~\ref{sec:proofs}). With Theorem~\ref{thm:inverse_fisher_ZI}, we have $\sqrt{n}(\hat\phi - \phi)  \stackrel{{\cal L}}{\rightarrow} N\left(0, \phi (1-\phi) \frac{d_{\boldsymbol\theta} \delta_{\boldsymbol\theta}}{d_{\boldsymbol\theta}\delta_{\boldsymbol\theta} - p_0(\boldsymbol\theta)}\right)$ and $\sqrt{n}(\hat{\boldsymbol\theta} - \boldsymbol\theta)  \stackrel{{\cal L}}{\rightarrow} N\left({\mathbf 0}, {\mathbf D}_{\boldsymbol\theta} \right)$. The relevant confidence intervals and hypothesis tests can be performed similarly as for zero-altered models. 

As mentioned in Section~\ref{sec:ZAP}, we need to calculating MLE and Fisher information accurately and efficiently. Explicit formulae of the Fisher information matrix of zero-inflated Poisson (ZIP) has been provided in Example~5 of \cite{aldirawi2022modeling}. In this section, we provide explicit formulae of the gradients and Fisher information matrices for zero-inflated geometric (ZIGe) and zero-inflated negative binomial (ZINB) models. We relegate the corresponding formulae for zero-inflated beta binomial (ZIBB) and zero-inflated beta negative binomial (ZIBNB) models to the Supplementary Materials (Section~\ref{sec:more_examples}).

\begin{example}\label{ex:ZIGe}{\bf Zero-inflated geometric model (ZIGe)} {\rm 
Same as in Example~\ref{ex:ZAGEOM}, the pmf of the baseline distribution can be written as $f_{p} (y) = p(1-p)^{y}$ with $y \in \{0, 1, 2, \ldots\}$ and parameter $p\in (0,1)$. 
According to Theorem~5 in \cite{aldirawi2022modeling}, the Fisher information matrix of the corresponding ZIGe distribution is
\begin{eqnarray*}
{\mathbf F}_{\rm ZIGe}= \frac{1}{\phi + (1-\phi)p} \begin{bmatrix}
\frac{1-p}{1-\phi} & 1 \\
1 & \frac{(1-\phi)[\phi(1-p) + p(1-\phi) + \phi p^2]}{p^2 (1-p)}
\end{bmatrix}
\end{eqnarray*}
\hfill{$\Box$}
}
\end{example}

\begin{example}\label{ex:ZINB}{\bf Zero-inflated negative binomial model (ZINB)} {\rm
Different from Example~3 of \cite{aldirawi2022modeling}, we take the form of the pmf of the baseline NB distribution as $f_{\boldsymbol\theta}(y)= \frac{\Gamma(y+r)}{\Gamma(y+1) \Gamma(r)} p^{r}(1-p)^{y}$ with parameters $\boldsymbol\theta = (r, p) \in (0, \infty)\times [0,1]$, $y \in \{0, 1, 2, \ldots\}$, which is more popular in the statistical literature (see, for example, \cite{hogg2019introduction}). In short, the $p$ in Example~3 of \cite{aldirawi2022modeling} is replaced by $1-p$ here. Then $p_0(\boldsymbol\theta) = p^r$ and 
\begin{eqnarray*}
	\frac{\partial\log f_{\boldsymbol\theta}(y)}{\partial r} &=& \Psi(y+r)-\Psi(r)+\log p\\
	\frac{\partial\log f_{\boldsymbol\theta}(y)}{\partial p} &=& \frac{r}{p}-\frac{y}{1-p}\\
	\frac{\partial\log p_0({\theta})}{\partial r} &=& \log p\\
	\frac{\partial\log p_0({\theta})}{\partial p} &=& \frac{r}{p}
\end{eqnarray*}
where $\Psi(\cdot)$ is the digamma function. According to Theorem~5 in \cite{aldirawi2022modeling}, the Fisher information matrix of the ZINB distribution is 
\[
{\mathbf F}_{\rm ZINB} = \left[
\begin{array}{cc}
A_{11} & A_{12} \>\>\>\> A_{13} \\
\begin{array}{c} A_{12}\\ A_{13}\end{array} & {\mathbf F}_{{\rm ZINB}\boldsymbol\theta}
\end{array}\right] 
\]
where $A_{11} = \frac{1-p^r}{[\phi+(1-\phi)p^r](1-\phi)}$, $A_{12} = \frac{p^r \log p}{\phi+(1-\phi)p^r}$,  $A_{13} = \frac{r p^{r-1}}{\phi+(1-\phi)p^r}$, and 
\[
{\mathbf F}_{{\rm ZINB}\boldsymbol\theta} = -(1-\phi) \left(\begin{bmatrix}
B_{11} & B_{12} \\
B_{12} & B_{22} 
\end{bmatrix}+
\frac{\phi p^r}{\phi+(1-\phi)p^r}
\begin{bmatrix}
C_{11} & C_{12} \\
C_{12} & C_{22} 
\end{bmatrix}\right)
\]
with $B_{11} = E\Psi_{1}(Y'+r)-\Psi_{1}(r)$, $B_{12} = \frac{1}{p}$, $B_{22} = -\frac{r}{p^2(1-p)}$, $C_{11} = (\log p)^2$, $C_{12} = \frac{r \log p}{p}$, and $C_{22} = \frac{r^2}{p^2}$. Here $\Psi_1(\cdot)$ is the trigamma function (see Example~\ref{ex:ZABB}).
}
\hfill{$\Box$}
\end{example}



\subsection{Zero-altered and zero-inflated models with continuous baseline distributions}\label{sec:za_zi}

As mentioned in Section~\ref{sec:zimle}, if the baseline distribution $f_{\boldsymbol\theta}(y)$ satisfies $p_0(\boldsymbol\theta) =  P_{\boldsymbol\theta}(Y=0) = 0$ given $Y\sim f_{\boldsymbol\theta}(y)$, then the zero-altered model~\eqref{eq:hurdle} and the zero-inflated model~\eqref{eq:zimodel} are the same. We call such kind of models the {\it zero-altered-zero-inflated} (ZAZI) models. Examples include all continuous baseline distributions, as well as discrete or mixture baseline distributions satisfying $p_0(\boldsymbol\theta)=0$. For ZAZI models with a baseline distribution $f_{\boldsymbol\theta}(y)$, its distribution function can be written as
\begin{equation}\label{eq:zazi_model}
f_{\rm ZAZI}(y\mid \phi, \boldsymbol\theta) = \phi {\mathbf 1}_{\{y=0\}} + (1-\phi) f_{\boldsymbol\theta}(y) {\mathbf 1}_{\{y\neq 0\}}
\end{equation}
Given a random sample $Y_1, \ldots, Y_n \sim f_{\rm ZAZI}(y\mid \phi, \boldsymbol\theta)$, the likelihood function of $(\phi, \boldsymbol\theta)$ is 
\begin{equation}\label{eq:zazi_like}
	L(\phi, \boldsymbol\theta) = \phi^{n-m} (1-\phi)^m \cdot \prod_{i: Y_i \neq 0} f_{\boldsymbol\theta}(Y_i)
\end{equation}
where $m = \#\{i: Y_i\neq 0\}$.	Then the MLEs maximizing \eqref{eq:zazi_like} are $\hat\phi = 1-m/n$ and $\hat{\boldsymbol\theta} = {\rm argmax}_{\boldsymbol\theta} \prod_{i: Y_i \neq 0} f_{\boldsymbol\theta}(Y_i)$, which are similar as \eqref{eq:hurdlemle} for zero-altered models. As a special case of Theorem~3 in \cite{aldirawi2022modeling}, we have the following formulae for the Fisher information matrix of ZAZI distributions.
	
\begin{theorem}\label{thm:continous FI}{\rm
Under regularity conditions, the Fisher information matrix of the ZAZI distribution~\eqref{eq:zazi_model} is \[
{\mathbf F}_{\rm ZAZI} = \begin{bmatrix}
\phi^{-1}(1-\phi)^{-1} & {\mathbf 0}^{T}\\
{\mathbf 0} & {\mathbf F}_{{\rm ZAZI}\boldsymbol\theta}
\end{bmatrix}
\]
where 
\[
{\mathbf F}_{{\rm ZAZI}\boldsymbol\theta} = - (1-\phi)\cdot E\left[\frac{\partial^{2}\log f_{\boldsymbol\theta}(Y')}{\partial \boldsymbol\theta \partial \boldsymbol\theta^{T}}\right]
\]
and $Y'$ follows the baseline distribution $f_{\boldsymbol\theta}(y)$.
\hfill{$\Box$}
}\end{theorem}

In this section, we provide explicit formulae of gradients and Fisher information matrices for commonly used ZAZI models with continuous baseline distributions including Gaussian or normal, log-normal, half-normal, and exponential distributions.

\begin{example}\label{ex:ZA/ZI normal}{\bf Zero-altered-zero-inflated Gaussian model (ZAZIG)} {\rm
This model has been used by, for example, \cite{zhang2020zero} for analyzing longitudinal microbiome data, known as a zero-inflated Gaussian (ZIG) model. The pdf of the baseline distribution with parameters $\theta=(\mu,\sigma)\in \mathbb{R}\times(0,\infty)$ is given by $f_{\boldsymbol\theta}(y)= \frac{1}{\sqrt{2\pi\sigma^{2}}}\exp\{-\frac{1}{2\sigma^{2}}(y-\mu)^{2}\} $. Then 
\begin{eqnarray*}
\frac{\partial\log f_{\boldsymbol\theta}(y)}{\partial \mu} &=& \frac{y-\mu}{\sigma^{2}} \\
\frac{\partial\log f_{\theta}(y)}{\partial \sigma}&=& - \frac{1}{\sigma} + 
\frac{(y- \mu)^{2}}{\sigma^{3}}
\end{eqnarray*}
According to Theorem~\ref{thm:continous FI}, the Fisher information matrix of the ZAZIG distribution is 
\[
{\mathbf F}_{\rm ZAZIG} =\begin{bmatrix}
\phi^{-1}(1-\phi)^{-1} & 0 & 0\\
0 & \frac{1-\phi}{\sigma^2} & 0\\
0 & 0 & \frac{2(1-\phi)}{\sigma^2}
\end{bmatrix}
\]
Note that in this case, given a random sample $Y_1, \ldots, Y_n$ from ZAZIG, $\hat\mu = \frac{1}{m} \sum_{i:Y_i\neq 0} Y_i$ and $\hat\sigma = \left(\frac{1}{m} \sum_{i:Y_i\neq 0} (Y_i - \hat\mu)^2\right)^{1/2}$ have explicit formulae, where $m = \#\{i: Y_i\neq 0\}$. 
}\hfill{$\Box$}
\end{example}

\begin{example}\label{ex:ZA/ZI log-normal}{\bf Zero-altered-zero-inflated log-normal model (ZAZILN)}  {\rm
This model, also known as zero-inflated log-normal (ZILN) model, has been used by, for example, \cite{zhou1999comparison} to study the effects of a prospective DUR intervention program for randomized clinical trials. The pdf of the baseline distribution with parameters $\boldsymbol\theta=(\mu,\sigma)\in \mathbb{R}\times(0,\infty)$ is given by $f_{\boldsymbol\theta}(y)=
\frac{1}{y\sqrt{2\pi\sigma^{2}}}\exp\{-\frac{1}{2\sigma^{2}}(\log y-\mu)^{2}\}$, $y>0$. Then 
\begin{eqnarray*}
\frac{\partial\log f_{\boldsymbol\theta}(y)}{\partial \mu} &=& \frac{\log y-\mu}{\sigma^{2}} \\
\frac{\partial\log f_{\boldsymbol\theta}(y)}{\partial \sigma} &=& - \frac{1}{\sigma} + 
\frac{(\log y- \mu)^{2}}{\sigma^{3}}
\end{eqnarray*}
According to Theorem~\ref{thm:continous FI}, the Fisher information matrix of the ZAZILN distribution is 
\[
{\mathbf F}_{\rm ZAZILN} =\begin{bmatrix}
\phi^{-1}(1-\phi)^{-1} & 0 & 0\\
0 & \frac{1-\phi}{\sigma^2} & 0\\
0 & 0 & \frac{2(1-\phi)}{\sigma^2}
\end{bmatrix}
\]
which is exactly the same as the one in Example~\ref{ex:ZA/ZI normal}.
\hfill{$\Box$}
}
\end{example}

\begin{example}\label{ex:ZA/ZI halfnormal}{\bf Zero-altered-zero-inflated half-normal model (ZAZIHN)}  {\rm
Also known as a zero-inflated half-normal model (ZIHN), this model has been used by, for example, \cite{reyna2012searching} as a candidate distribution for modeling the animal movement distances in the wild. In our notations, the pdf of its baseline distribution with parameter $\boldsymbol\theta = \sigma \in (0,\infty)$, known as a standard half-normal distribution, is given by $f_{\boldsymbol\theta}(y)=
\frac{\sqrt{2}}{\sqrt{\pi\sigma^{2}}}\exp\{-\frac{y^{2}}{2\sigma^{2}}\}$, $y>0$.
Then 
$\frac{\partial\log f_{\boldsymbol\theta}(y)}{\partial \sigma} = - \frac{1}{\sigma} + \frac{y^{2}}{\sigma^{3}}$. According to Theorem~\ref{thm:continous FI}, the Fisher information matrix of the ZAZIHN distribution is 
\[
{\mathbf F}_{\rm ZAZIHN} = \begin{bmatrix}
\phi^{-1}(1-\phi)^{-1} & 0\\
0 & \frac{2(1-\phi)}{\sigma^{2}}
\end{bmatrix}
\]
It should be noted that other forms of ZIHN models have also been used in the literature. For example, \cite{chen2003random} utilized a more general ZIHN distribution as a prior for a hierarchical Bayesian model. The continuous component of their ZIHN is a general normal distribution $N(\mu, \sigma^2)$ truncated below
by zero. In that case, $\boldsymbol\theta = (\mu, \sigma)$.
\hfill{$\Box$}
}
\end{example}

\begin{example}\label{ex:ZA/ZI exponential}{\bf Zero-altered-zero-inflated exponential model (ZAZIE)} {\rm
Also known as zero-inflated exponential model (ZIE), this model has been used by, for example, \cite{huang2019zero} for modeling casualty rates in ship collision. The pdf of its baseline distribution with parameter $\boldsymbol\theta = \lambda \in (0,\infty)$ can be written as $f_{\boldsymbol\theta}(y)=
\lambda e^{-\lambda y}$, $y > 0$. Then $\frac{\partial\log f_{\boldsymbol\theta}(y)}{\partial \lambda} = \frac{1}{\lambda} - y$. According to Theorem~\ref{thm:continous FI}, the Fisher information matrix of the ZAZIE distribution is 
\[
{\mathbf F}_{\rm ZAZIE} = \begin{bmatrix}
\phi^{-1}(1-\phi)^{-1} & 0\\
0 & \frac{1-\phi}{\lambda^{2}}
\end{bmatrix}
\]
Note that the MLE of $\lambda$ has an explicit form $\hat\lambda = \frac{m}{\sum_{i:Y_i\neq 0} Y_i}$ given a random sample $Y_1, \ldots, Y_n$ from ZAZIE, where $m = \#\{i: Y_i\neq 0\}$. 
\hfill{$\Box$}
}
\end{example}

\subsection{Model selection based on KS and likelihood ratio tests}\label{sec:model_selection}

Given so many zero-altered or zero-inflated models listed in Sections~\ref{sec:ZAP}, \ref{sec:zimle} and \ref{sec:za_zi}, a critical question is which model is the most appropriate one for a given dataset.

The Kolmogorov-Smirnov (KS) test has been commonly used for testing whether a random sample $\{Y_1, \ldots, Y_n\}$ comes from a continuous cumulative distribution function $F_{\boldsymbol\theta}(y)$ with specified model parameter(s) $\boldsymbol\theta$ \citep{massey1951kolmogorov}. It is based on the KS statistic $D_n = \sup_{y} |F_n(y) - F_{\boldsymbol\theta}(y)|$, where $F_n(y) = n^{-1} \sum_{i=1}^n {\mathbf 1}_{(-\infty, y]}(Y_i)$ is known as the {\it empirical distribution function}. \cite{dimitrova2020computing} extended the KS test for general distributions $F_{\boldsymbol\theta}(y)$ with known parameter(s) $\boldsymbol\theta$, including discrete and mixed ones. For typical applications, $\boldsymbol\theta$ is unknown and an estimate $\hat{\boldsymbol\theta}$ is plugged in when calculating $D_n$, which tends to overestimate the corresponding $p$-value \citep{lilliefors1967kolmogorov, lilliefors1969kolmogorov, aldirawi2019identifying}. To overcome the biasedness of estimated $p$-value due to the plugged-in estimated parameters, \cite{aldirawi2019identifying} proposed a bootstrapped Monte Carlo estimate $ \frac{\#\{b\mid D_n^{(b)} > D_n\} + 1}{B+1}$ for the $p$-value in their Algorithm~1, where $D_n^{(b)} = \sup_y |{F}_n^{(c)}(y) - F_{\hat{\boldsymbol\theta}^{(b)}}(y)|$, ${F}_n^{(c)}(y)$ is the empirical distribution function of a random sample ${\mathbf Y}^{(c)} = \{Y_1^{(c)}, \ldots, Y_n^{(c)}\}$ from $F_{\hat{\boldsymbol\theta}^{(b)}}$, $\hat{\boldsymbol\theta}^{(b)}$ is the MLE based on a bootstrapped sample ${\mathbf Y}^{(b)} = \{Y_1^{(b)}, \ldots, Y_n^{(b)}\}$ of $\{Y_1, \ldots, Y_n\}$, $b=1, \ldots, B$, and $B$ is a predetermined large number, typically $B=1000$. If the estimated $p$-value is larger than $0.05$, we say that the specified distribution passes the KS test for the given dataset. 

Since $D_n = \sup_y |F_n(y) - F_{\hat{\boldsymbol\theta}}(y)|$ when $\boldsymbol\theta$ is unknown, where both $F_n$ and $\hat{\boldsymbol\theta}$ are based on the same data $\{Y_1, \ldots, Y_n\}$, a more reasonable bootstrapped version of $D_n$ would be $D_n^{(b)'} = \sup_y |{F}_n^{(c)}(y) - F_{\hat{\boldsymbol\theta}^{(c)}}(y)|$, where $\hat{\boldsymbol\theta}^{(c)}$ is the MLE based on ${\mathbf Y}^{(c)} = \{Y_1^{(c)}, \ldots, Y_n^{(c)}\}$ instead of ${\mathbf Y}^{(b)}$. In this paper, we propose the following nested bootstrap estimate for KS test $p$-value based on the above argument.

\begin{algorithm}\caption{Nested Bootstrap Algorithm for Estimating  $p$-value of KS Test}\label{algo:modified_KS_test}
\begin{itemize}
		\item[1:] Given data ${\mathbf Y}=\{Y_1, Y_2, \cdots Y_n\}$, calculate the MLE $\hat{\boldsymbol\theta}$ of $\boldsymbol\theta$ and the KS statistic $D_{n} = \sup_{y}|  F_{n}(y)-  F_{\hat{\boldsymbol\theta}}(y)|$.
		\item[2:] For $b = 1, \ldots, B$, do steps 3$\sim$7.
        \item[3:] Resample ${\mathbf Y}$ with replacement to get a bootstrapped sample ${\mathbf Y}^{(b)} = \{Y^{(b)}_{1}, \cdots, Y_{n}^{(b)}\}$. 
		\item[4:] Calculate the MLE $\hat{\boldsymbol\theta}^{(b)}$ of $\boldsymbol\theta$ based on ${\mathbf Y}^{(b)}$.
		\item[5:] Simulate a random sample ${\mathbf Y}^{(c)} = \{Y^{(c)}_1, \ldots, Y^{(c)}_n\}$ from $F_{\hat{\boldsymbol\theta}^{(b)}}$. 
		\item[{ 6:}]  Calculate the MLE $\hat{\boldsymbol\theta}^{(c)}$ of $\boldsymbol\theta$ based on ${\mathbf Y}^{(c)}$. 
		\item[{ 7:}] { Calculate the bootstrapped KS statistic $D_{n}^{(b)'}={\rm sup}_{y}|  F^{(c)}_{n}(y)- F_{\hat\theta^{(c)}}(y)|$}, where $F^{(c)}_{n}(y)$ is the empirical distribution function of ${\mathbf Y}^{(c)}$. 
		\item [8:] Estimate the $p$-value of the KS test by 
		$\frac{\#\{b \mid D_{n}^{(b)'} > D_{n}\}}{B}$~.
	\end{itemize}
\end{algorithm}

To make a distinction, we call \cite{aldirawi2019identifying}'s Algorithm~1 as Algorithm~1A and our Algorithm~\ref{algo:modified_KS_test} as Algorithm~1B. The corresponding {\tt R} functions are named {\tt kstest.A} and {\tt kstest.B}, respectively. According to our simulation studies in Section~\ref{sec:simulation_study_Algorithm1}, we recommend {\tt kstest.B} for small sample sizes such as $n=30,50$. For larger sample sizes, since the difference in terms of test power is negligible, we recommend {\tt kstest.A} for less computational cost. 

In practice, it is not uncommon that two or more distributions pass the KS test for the same dataset, especially when the sample size is moderate or small (see, for example, \cite{aldirawi2019identifying}). In this situation, likelihood ratio tests may be used for pairwise comparisons. More specifically, for testing $H_0: Y_1, \cdots, Y_n$ iid $\sim f(y; \boldsymbol\theta)$ with unknown parameter(s) $\boldsymbol\theta$ against	$H_1: Y_1, \cdots, Y_n$ iid $\sim g(y; \boldsymbol\delta)$ with unknown parameter(s) $\boldsymbol\delta$, the likelihood ratio test statistic in log scale (see, for example, \cite{aldirawi2019identifying}) can be defined as
\[
	\Lambda = \log \frac{\prod_{i=1}^{n}f(Y_i; \hat{\boldsymbol\theta})}{\prod_{i=1}^{n}g(Y_i; \hat{\boldsymbol\delta})}
\]
where $\hat{\boldsymbol\theta}$ and $\hat{\boldsymbol\delta}$ are the corresponding maximum likelihood estimates. Smaller $\Lambda$ values are in favor of the alternative distribution $g(y; \boldsymbol\delta)$. 
	
To overcome the possible biasedness due to plugged-in estimated parameters, we adopt the bootstrapped estimate of $p$-value described by Algorithm~2 of \cite{aldirawi2019identifying}.
More specifically, (i) bootstrap samples ${\mathbf Y}^{(b)} = \{Y_1^{(b)}, \ldots, Y_n^{(b)}\}$, $b=1, \ldots, B$ are obtained from the original data ${\mathbf Y} = \{Y_1, \ldots, Y_n\}$; (ii) MLEs $\hat{\boldsymbol\theta}^{(b)}$ and $\hat{\boldsymbol\delta}^{(b)}$ are calculated based on ${\mathbf Y}^{(b)}$; (iii) a random sample ${\mathbf Y}^{(c)} = \{Y_1^{(c)}, \ldots, Y_n^{(c)}\}$ is simulated from $f(y; \hat{\boldsymbol\theta}^{(b)})$; (iv) bootstrapped test statistic $\Lambda^{(b)}$ is calculated based on ${\mathbf Y}^{(c)}$; and (v) the estimated $p$-value is $\frac{\#\{b\mid \Lambda^{(b)} < \Lambda\}}{B}$. If the $p$-value is less than $0.05$, we claim that $H_1$ is significantly better; otherwise, we stick to $H_0$.

\subsection{Zero-altered model versus zero-inflated model}\label{sec:za_vs_zi}

Clearly a zero-altered model~\eqref{eq:hurdle} and its corresponding  zero-inflated model~\eqref{eq:zimodel} are connected by sharing the same baseline distribution $f_{\boldsymbol\theta}(y)$. Under some conditions, they are actually equivalent due to the following theorem.

\begin{theorem}\label{thm:za_vs_zi} Let $f_{\rm ZA}(y\mid \phi_{\rm ZA}, \boldsymbol\theta)$ be a zero-altered model as in \eqref{eq:hurdle} and $f_{\rm ZI}(y\mid \phi_{\rm ZI}, \boldsymbol\theta)$ be the corresponding zero-inflated model as in \eqref{eq:zimodel} with the same baseline distribution $f_{\boldsymbol\theta}(y)$.  
\begin{itemize}
    \item[(i)] Given $f_{\rm ZI}(y\mid \phi_{\rm ZI}, \boldsymbol\theta)$, we let $\phi_{\rm ZA} = \phi_{\rm ZI} + (1-\phi_{\rm ZI}) p_0(\boldsymbol\theta)$ and then $f_{\rm ZA}(y\mid \phi_{\rm ZA}, \boldsymbol\theta) = f_{\rm ZI}(y\mid \phi_{\rm ZI}, \boldsymbol\theta)$ for all $y$.
    \item[(ii)] Given $f_{\rm ZA}(y\mid \phi_{\rm ZA}, \boldsymbol\theta)$ with $\phi_{\rm ZA} \geq p_0(\boldsymbol\theta)$, we let $\phi_{\rm ZI} = [\phi_{\rm ZA} - p_0(\boldsymbol\theta)]/[1- p_0(\boldsymbol\theta)]$ and then $f_{\rm ZI}(y\mid \phi_{\rm ZI}, \boldsymbol\theta) = f_{\rm ZA}(y\mid \phi_{\rm ZA}, \boldsymbol\theta)$ for all $y$.
\end{itemize}
\end{theorem}

The proof of Theorem~\ref{thm:za_vs_zi} is relegated to the Supplementary Materials (Section~\ref{sec:proofs}). Theorem~\ref{thm:za_vs_zi} implies that if $\phi_{\rm ZA} \geq p_0(\boldsymbol\theta)$, which indicates that the data is zero-inflated, then a ZA model is equivalent to the corresponding ZI model.
Given a random sample from a zero-inflated model $f_{\rm ZI}(y\mid \phi_{\rm ZI}, \boldsymbol\theta)$, a KS test will conclude that some zero-altered model $f_{\rm ZA}(y\mid \phi_{\rm ZA}, \boldsymbol\theta)$ seems fine with the data as well. The other direction is slightly different though. That is, if the true model is a zero-altered one, only when $\phi_{\rm ZA}$ is at least as large as $p_0(\boldsymbol\theta)$, a KS test will conclude that some zero-inflated model seems to be true as well. In Section~\ref{sec:realdata}, we will provide a numerical example such that both ZIBNB and BNBH fit the data well.

\section{Software Implementation and Numerical Analysis}\label{sec:numerical_study}

In this section, the {\tt R} implementation of the proposed package {\tt AZIZD} is introduced and its numerical analysis is performed. Real data examples are used to illustrate how our package can be used in practice.

Compared with existing {\tt R} packages each covering only a limited number of baseline distributions, our package covers many more discrete and continuous distributions including Poisson, geometric, negative binomial, beta binomial, beta negative binomial, normal (or Gaussian), log-normal, half-normal, and exponential distributions along with their zero-altered and zero-inflated models, which facilitates the potential users to choose the most appropriate model from a large class of candidates for their dataset. For all the models mentioned above, we provide the corresponding Fisher information matrix and confidence intervals for estimated parameters, which allows users to run hypothesis tests and make further inference.

\subsection{Improvements over existing packages on finding MLEs}\label{sec:algo_inflated_mle}	

From Section~\ref{sec:model_selection}, we can see that an accurate MLE is a critical component for both KS test and likelihood ratio test. In this section, we first summarize the improvements of the proposed {\tt AZIAD} over existing {\tt R} packages on finding MLEs for zero-altered and zero-inflated models.

\begin{itemize}
\item[(1)] Compared with other packages, {\tt AZIAD} provides MLEs well even for extreme cases including $\phi=0$ or $\phi=1$ (see the comparison analysis in Section~\ref{sec:simulation_study_Algorithm1_type_I}).
\item[(2)] Some packages encountered error message {\tt L-BFGS-B needs finite values of "fn"} for some zero-inflated data when using, for example, the function {\tt dis.kstest} in package {\tt iZID}. This issue is solved in {\tt AZIAD} by specifying appropriate lower bound and upper bound for optimizations (Section~\ref{sec:simulation_study_Algorithm1_type_I}).
\item[(3)] Compared with package {\tt iZID} which also covered MLEs for ZIBNB, BNBH, ZIBB, and BBH models, our results are more reliable (see Example~\ref{ex:zih.mlelsimulation} for a comparison) by applying Theorem~4 in \cite{aldirawi2022modeling}. More specifically, we separate the two situations (1) $m/n \leq 1 - p_0(\boldsymbol\theta_*)$ and (2) $m/n > 1-p_0(\boldsymbol\theta_*)$ and calculate the MLEs in each case.
\item[(4)] Some parameters are originally defined as positive integers, such as $n$ in Examples~\ref{ex:ZABB} and \ref{ex:ZIBB} and $r$ in Examples~\ref{ex:ZABNB} and \ref{ex:ZINB}, while in practice they could be extended to positive real numbers. In {\tt AZIAD}, we seek for real-valued MLEs by default. In the mean time, we keep the option of integer-valued $n$ or $r$ in response to users' call.
\end{itemize}

In the rest part of this section, we use examples to illustrate how to use our package to find the MLEs and the improved accuracy by using our package. 

\begin{example}\label{ex:zih.mlel} {\rm
In order to find the MLE for the parameter(s) of zero-inflated or hurdle models, the main function built in {\tt AZIAD} is 
\begin{verbatim}
zih.mle(x, r, p, alpha1, alpha2, n, lambda, mean, sigma, type = c("zi", 
"h"), dist, lowerbound = 0.01, upperbound = 10000)
\end{verbatim}
where {\tt x} is a sequence of numbers, which could be integers for discrete cases or real numbers for continuous cases; the arguments {\tt r}, {\tt p}, {\tt alpha1}, {\tt alpha2}, {\tt n}, {\tt lambda}, {\tt mean}, and {\tt sigma} are initial values of the corresponding parameters;  {\tt dist} could be chosen as {\tt poisson.zihmle}, {\tt geometric.zihmle}, {\tt nb.zihmle}, {\tt nb1.zihmle}, {\tt bb.zihmle}, {\tt bb1.zihmle}, {\tt bnb.zihmle}, {\tt bnb1.zihmle}, {\tt normal.zihmle}, {\tt halfnorm.zihmle}, {\tt lognorm.zimle}, and {\tt exp.zihmle}, which correspond to zero-inflated or zero-altered Poisson, geometric, negative binomial, negative binomial with integer-valued $r$, beta binomial, beta binomial with integer-valued $n$, beta negative binomial, beta negative binomial with integer-valued $r$, normal, log-normal, half-normal, and exponential distributions, respectively; option {\tt type} indicates the type of distribution is zero-inflated ({\tt zi}) or hurdle/zero-altered ({\tt h}); {\tt lowerbound} and {\tt upperbound} specify the searching range of parameters when maximizing the likelihood function. 
For instance, in order to calculate the MLE of zero-inflated geometric distribution (see Example~\ref{ex:ZIGe}), one may use the following {\tt R} code:
\begin{verbatim}
R> set.seed(008)
R> x22=sample.h1(2000,phi=0.3,dist='geometric',p=0.3)
R> zih.mle(x22,p=0.2,dist="geometric.zihmle",type="h")               
             p    phi   loglik
      0.2942292 0.3015 -4101.05
\end{verbatim}
Our estimates $\hat{p} = 0.2942292$ and $\hat\phi = 0.3015$ are fairly close to the true parameter values $p=0.3$ and $\phi=0.3$.}\hfill{$\Box$}
\end{example}

\begin{example}\label{ex:zih.mlelsimulation} {\rm
To compare the performance of our {\tt AZIAD} package and the existing {\tt iZID} on finding MLEs, we generate random samples from a BNBH model (see Example~\ref{ex:ZABNB}) with parameters $(\phi,r,\alpha_1,\alpha_2) = (0.3,5,8,3)$, 
and a BBH model (see Example~\ref{ex:ZABB}) with parameters $(\phi,n,\alpha_1,\alpha_2) = (0.6,5,8,3)$, each with increasing sample sizes $N=10^4, 5\times 10^4, 20\times 10^4$ and $100\times 10^4$. To see the converging pattern more clearly, we generate nested datasets such that any dataset with a smaller $N$ is a subset of the corresponding datasets with bigger $N$s, if they are simulated from the same distribution.

For the each sample, we find the MLEs $(\hat\phi, \hat\theta_1, \hat\theta_2, \hat\theta_3)$ of the parameters and calculate the aggregated $L_1$ relative distance (L1RD)  $L_1 = |\hat\phi - \phi|/|\phi| + \sum_{i=1}^3 |\hat\theta_i - \theta_i|/|\theta_i|$. The implemented {\tt R} codes are as follows:

\begin{verbatim}
R> set.seed(167)
R> hi1=sample.h1(N=1000000,phi=0.3,dist="bnb",r=5,alpha1=8,alpha2=3)
R> hi2=hi1[1:200000]
R> hi3=hi1[1:50000]
R> hi4=hi1[1:10000]
R> mle13=zih.mle(hi4,type="h",r=6,alpha1=9,alpha2=4,dist="bnb.zihmle")
R> mle13=zih.mle(hi4,type="h",r=6,alpha1=9,alpha2=4,dist="bnb1.zihmle")
R> mle14=bnb.zihmle(hi4,type="h",r=6,alpha1=9,alpha2=4)
\end{verbatim}

\begin{verbatim}
R> set.seed(171)
R> hi1=sample.h1(N=1000000,phi=0.6,dist="bb",n=5,alpha1 = 8,alpha2=3)
R> hi2=hi1[1:200000]
R> hi3=hi1[1:50000]
R> hi4=hi1[1:10000]
R> mle23=zih.mle(hi1,n=6,alpha1=9,alpha2=4,type="h",dist="bb.zihmle")
R> mle23=zih.mle(hi4,n=6,alpha1=9,alpha2=4,type="h",dist="bb1.zihmle")
R> mle24=bb.zihmle(hi1,n=6,alpha1=9,alpha2=3,type="h")
\end{verbatim}

\begin{table}
	\centering
		\begin{tabular}{@{}lllllllllllll@{}}
			\hline
			$N$ & {\tt R} function & $r$ & $\alpha_{1}$ & $\alpha_{2}$ & $\phi$ & loglike &  L1RD \\ \midrule
			$1\times 10^4$ & zih.mle(bnb) & 3.81 & 7.69 & 3.81 & 0.303 & -19232.34 & 0.555 \\  
			&zih.mle(bnb1) & 4 & 7.70 & 3.64 & 0.303 & -19232.35& 0.460\\
			& bnb.zihmle & 5.83 & 20.68 & 3.89 & 0 & 777027.7 & 3.05 \\ \midrule
			$5\times 10^4$ & zih.mle(bnb) & 5.35 & 7.69 & 2.67 & 0.301 & -96201.68 & 0.223 \\  
			&zih.mle(bnb1) & 5 & 7.59 & 2.82 & 0.301 & -96201.71 & 0.113\\
			& bnb.zihmle & 30.09 & 453.25 & 28.96 & 0 & 134618737 & 70.32 \\ \midrule
			$20\times 10^4$ & zih.mle(bnb) & 5.36 & 7.88 & 2.74 & 0.299 & -384542.1 & 0.175 \\
			&zih.mle(bnb1) & 5 & 7.78 & 2.90 & 0.299& -384542.2 & 0.061\\
			& bnb.zihmle & 92.90 & 5492.33 & 119.79 & 0 & 8730091135 & 743.05 \\
			\midrule
			$1\times 10^6$ & zih.mle(bnb) & 5.46 & 8.01 & 2.78 & 0.299 & -1921040 & 0.181 \\
			&zih.mle(bnb1) & 5 & 7.97 & 2.98 & 0.299 & -1921041 & 0.009\\
			& bnb.zihmle & 101.23 & 6842.71 & 138.64 & 0 & 55719787227 & 919.80 \\
			\midrule
		\end{tabular}
		\caption{Comparison of MLE estimates, log-likelihood, and $L_1$ relative distance (L1RD) for BNB Hurdle distribution with true parameters ($r=5$, $\alpha_1=8$, $\alpha_2=3$, $\phi=0.3$) using {\tt bnb.zihmle} in {\tt iZID} and {\tt zih.mle} in {\tt AZIAD} with
		{\tt dist=bnb.zihmle} (real-valued $r$) and {\tt dist=bnb1.zihmle} (integer-valued $r$) with various sample sizes $N$}
		\label{table:BNBH}
\end{table}

\begin{table}
	\centering
		\begin{tabular}{@{}lllllllllllll@{}}
			\hline
			$N$ & {\tt R} function & $n$ & $\alpha_{1}$ & $\alpha_{2}$ & $\phi$ & loglike & L1RD \\ \midrule
			$1\times 10^4$ & zih.mle(bb) & 4.99 & 7.58 & 2.84 & 0.597 & -12572.96 & 0.110 \\ 
			& zih.mle(bb1) & 5 & 7.86 & 2.96 & 0.597 & -12579.31 & 0.033\\
			& bb.zihmle & 6 & 9 & 3 & 0.597 & 28775.56 & 0.33 \\ \midrule
			$5\times 10^4$ & zih.mle(bb) & 4.99 & 7.57& 2.78 & 0.597 & -62688.44 & 0.129 \\  
			& zih.mle(bb1) & 5 & 7.85 & 2.91 & 0.597& -62721.29 & 0.051\\
			& bb.zihmle & 6 & 9 & 3 & 0.597 & 143733.9 & 0.328 \\ \midrule
			$20\times 10^4$ & zih.mle(bb) & 4.99 & 7.67 & 2.84 & 0.600 & -249962 & 0.094 \\
		    & zih.mle(bb1) & 5 & 7.95 & 2.97 & 0.600 & -250089.3 & 0.015\\	
			& bb.zihmle & 6 & 9 & 3 & 0.600 & 570424.6 & 0.325  \\
			\midrule
			$1\times 10^6$ & zih.mle(bb) & 4.99 & 7.66 & 2.85 & 0.600 &  -1249976 & 0.093 \\
			& zih.mle(bb1) & 5 & 7.94 & 2.97 & 0.600 & -1250610 & 0.015\\
			& bnb.zihmle & 6 & 9 & 3 & 0.600 & 2850793 & 0.325 \\
			\midrule
		\end{tabular}
		\caption{Comparison of MLE estimates, log-likelihood, and $L_1$ relative distance (L1RD) for BB Hurdle distribution with true parameters ($n=5$, $\alpha_1=8$, $\alpha_2=3$, $\phi=0.6$) using {\tt bb.zihmle} in {\tt iZID} and {\tt zih.mle} in {\tt AZIAD} with
		{\tt dist=bb.zihmle} (real-valued $r$) and {\tt dist=bb1.zihmle} (integer-valued $r$) with various sample sizes $N$}
		\label{table:BBH}
\end{table}

In Tables~\ref{table:BNBH} and \ref{table:BBH}, we list and compare the MLEs, log-likelihood, and L1RD obtained by our package ({\tt R} function {\tt zih.mle} with real-valued or integer-valued MLEs) and the {\tt iZID} package ({\tt R} functions {\tt bnb.zihmle} and {\tt bb.zihmle}) for BNBH and BBH models.
We can see that the estimates based on our functions are more accurate as indicated by smaller L1RD. As the sample size increases, the L1RD based on our MLEs shows an overall decreasing pattern which indicates the convergence of MLEs towards the true parameter values. Similar results are collected for ZIBNB and ZIBB as well but not shown here.
}\hfill{$\Box$}
\end{example}

\subsection{Fisher information, confidence interval, and test on zero-inflation}\label{sec:fisher_ci}

As mentioned in Sections~\ref{sec:ZAP}, \ref{sec:zimle} and \ref{sec:za_zi}, the Fisher information matrix ${\mathbf F}_{\rm ZA}$, ${\mathbf F}_{\rm ZI}$ or ${\mathbf F}_{\rm ZAZI}$ can be calculated by {\tt AZIAD} for zero-altered/hurdle, zero-inflated, or ZAZI models, respectively. Its inverse matrix is approximately the variance-covariance matrix of parameter estimates $(\hat\phi, \hat{\boldsymbol\theta})$. 
For example, for zero-altered or hurdle models (see Section~\ref{sec:ZAP}), 
\[
\sqrt{n}(\hat\phi - \phi) \stackrel{\cdot}{\sim} N(0, \phi (1-\phi)), \>\>\>
\sqrt{n}(\hat{\boldsymbol\theta} - \boldsymbol\theta) \stackrel{\cdot}{\sim} N({\mathbf 0}, {\mathbf F}_{\rm ZA\boldsymbol\theta}^{-1})
\]
for large $n$. Approximate confidence intervals can be constructed for $\phi$ and $\boldsymbol\theta$ (for example, expression~\eqref{eq:hurdle_phi_ci} for $\phi$).

In our package {\tt AZIAD}, we use the function 
\begin{verbatim} 
FI.ZI(x, dist= "poisson", r = NULL, p = NULL, alpha1 = NULL,
alpha2 = NULL, n = NULL,lambda=NULL, mean=NULL, sigma=NULL,
lowerbound = 0.01, upperbound = 10000)
\end{verbatim}
to calculate (the inverse of) the Fisher information matrix at the MLE and 95\% approximate confidence intervals for all parameters, where {\tt x} is the data in its vector form; {\tt dist} can be {\tt poisson}, {\tt geometric}, {\tt nb}, {\tt bb}, {\tt bnb}, {\tt normal}, {\tt halfnormal}, {\tt lognormal}, {\tt exponential}, {\tt zip}, {\tt zigeom}, {\tt zinb}, {\tt zibb}, {\tt zibnb}, {\tt zinormal}, {\tt zilognorm}, {\tt zihalfnorm}, {\tt ziexp}, {\tt ph}, {\tt geomh}, {\tt nbh}, {\tt bbh}, and {\tt bnbh} for Poisson, geometric, negative binomial, beta binomial, beta negative binomial, normal/Gaussian, log-normal, half-normal, exponential, their zero-inflated versions and hurdle versions, respectively;
{\tt r}, {\tt p}, {\tt alpha1}, {\tt alpha2}, {\tt n}, {\tt lambda}, {\tt mean}, {\tt sigma} provide initial values of distribution parameters; {\tt lowerbound} and {\tt upperbound} are predetermined ranges for MLEs, which could be extended if attained by any of the estimated parameter values. 

\begin{example}\label{ex:fisherinformation} {\bf Confidence intervals for ZINB} {\rm\quad
As an illustration, we apply the {\tt FI.ZI} function to a simulated ZINB dataset with true parameters $\phi = 0.4$, $r=10$ and $p=0.2$. 

\begin{verbatim}
R> set.seed(117)
R> N=1000;r=10;p=0.2;phi=0.4;
R> x<-sample.zi1(N,phi=phi,dist="nb",r=r,p=p)
R> FI.ZI(x,r=3,p=0.1, dist="zinb")

$inversefisher
           [,1]       [,2]        [,3]
[1,] 0.32302489  0.0661767  0.02116832
[2,] 0.06617670 35.0781124 -0.53660185
[3,] 0.02116832 -0.5366019  0.03658915

$ConfidenceIntervals
               [,1]       [,2]
CI of phi 0.3937737  0.4642262
CI of r   9.8952803 10.6294496
CI of p   0.1923346  0.2160458
\end{verbatim}
With sample size $N=1000$, the derived 95\% confidence intervals cover the true parameter values fairly well.
\hfill{$\Box$}
}
\end{example}
	
	
\begin{example}\label{ex:ZIP_test_zero_inflation} {\bf Test zero-inflation in Poisson data} {\rm\quad
As an illustration, we simulate a ZIP data with $N=1000$, $\phi=0$ and $\lambda = 0.8$, which is actually a regular Poisson data with $\lambda = 0.8$ without zero-inflation. There are 459 zeros which roughly matches the probatility of zero $p_0(\lambda) = 0.449$. Then we use the {\tt FI.ZI} function with distribution Poisson Hurdle (PH) to allow both zero-inflation and deflation. 
\begin{verbatim}
R> set.seed(337)
R> x<-sample.zi1(N=1000,phi=0,dist="poisson",lambda=0.8)
R> table(x)
x
  0   1   2   3   4   5 
459 334 153  41  10   3 
R> dpois(0,lambda=0.8)
[1] 0.449329
R> FI.ZI(x,lambda=1, dist="ph")
$inversefisher
         [,1]     [,2]
[1,] 0.248319 0.000000
[2,] 0.000000 2.558941

$ConfidenceIntervals
                  [,1]      [,2]
CI of Phi    0.4281146 0.4898854
CI of lambda 0.7937409 0.9920343

R> FI.ZI(x,lambda=1, dist="poisson")
$inversefisher
lambda
[1,]  0.818

$ConfidenceIntervals
[1] 0.7619437 0.8740563
\end{verbatim}

It turns out that the 95\% approximate confidence interval $(0.4281, 0.4899)$ of $\phi$ based on PH model ({\tt dist="ph"}) covers the baseline zero probability $p_0(\lambda)=0.4493$, which indicates that neither zero-inflation nor zero-deflation is significant at the 5\% level. Actually, the 95\% approximate confidence interval $(0.7619, 0.8741)$ of $\lambda$ based on the regular Poisson model ({\tt dist="poisson"}) covers the true value $0.8$ roughly at the center, which is better than $(0.7937, 0.9920)$ based on the PH model.
}\hfill{$\Box$}
\end{example}

\subsection{Comparison study in terms of type I error
}\label{sec:simulation_study_Algorithm1_type_I}

In this section and Section~\ref{sec:simulation_study_Algorithm1}, we use simulation studies to compare the performance of our algorithms for model identification with existing {\tt R} functions for the same purposes. In this section we focus on type I errors. That is, given the true distribution, say ZINB, what is the chance that the test erroneously concludes that the distribution is not ZINB due to a $p$-value less than $0.05$. Such an error is known as a {\it type I error}~\citep{lehmann2005testing}. Ideally, such a chance is no more than $0.05$, known as the {\it size} of the test.

The {\tt R} functions under comparison in this section include the basic function {\tt ks.test}, function {\tt disc\_ks\_test} in package {\tt KSgeneral} \citep{dimitrova2020computing}, function {\tt dis.kstest} in package {\tt iZID} \citep{wang2020identifying}, and two of our functions in {\tt AZIAD}, {\tt kstest.A} based on \cite{aldirawi2019identifying}'s Algorithm~1 with $(\#b+1)/(B+1)$ replaced by $\# b/B$, and {\tt kstest.B} based on our Algorithm~\ref{algo:modified_KS_test} described in Section~\ref{sec:model_selection}.

For illustration purposes, we consider four zero-inflated models, ZIP, ZINB, ZIBNB, and ZIBB.
For each model, we consider five different samples sizes, $N=30, 50, 100, 200, 500$, and simulate $B=1000$ independent datasets for each sample size. For each simulated dataset, we run the five {\tt R} functions under comparison individually targeting the corresponding true model. If the $p$-value of the test is less than $0.05$, we count it as a type I error. The ratios of type I errors (that is, the number of type I errors divided by $B=1000$) are listed in Tables~\ref{tab:zip/zip}, \ref{tab:zinb/zinb}, \ref{tab:zibnb/zibnb} and \ref{tab:zibb/zibb}. For readers' reference, we provide the {\tt R} code for generating Table~\ref{tab:zinb/zinb} in the Supplementary Materials (Section~\ref{sec:more_codes}).

According to these tables, {\tt disc\_ks\_test} (package {\tt KSgeneral}) seems to have larger type I error rates than the nominal level $0.05$. 
One concern about function {\tt dis.kstest} ({\tt iZID}) for ZINB, ZIBNB and ZIBB distributions is its significant portions of {\tt NA}'s due to errors. Our two functions, {\tt kstest.A} and {\tt kstest.B}, and {\tt R} basic function {\tt ks.test} have type I error rates nearly zero, which are satisfactory.

\begin{table}[hbt!]
\centering
\begin{tabular}{@{}llllll@{}}
\toprule
    Sample Size $N$ & $30$ & $50$ & $100$ & $200$ & $500$
    \\ \midrule
ks.test & 0 & 0 & 0 & 0.001 & 0.005     \\ 
disc\_ks\_test & 0.074 & 0.116 & 0.081 & 0.112 & 0.08 \\
dis.kstest & 0 & 0 & 0 & 0 & 0    \\
kstest.A & 0 & 0 & 0.001 & 0 & 0     \\
kstest.B & 0 & 0 & 0 & 0 & 0      \\ \bottomrule
\end{tabular}
\caption{Type I error rates of KS tests on whether the data comes from ZIP, based on $B=1000$ simulated ZIP datasets with parameters $\phi=0.3$, $\lambda=10$ for each sample size $N$}
\label{tab:zip/zip}
\end{table}

\begin{table}[hbt!]
\centering
\footnotesize
\begin{tabular}{@{}llllll@{}}
\toprule
    Sample Size $N$ & $30$ & $50$ & $100$ & $200$ & $500$
    \\ \midrule
ks.test & 0.001 & 0.002 & 0.001 & 0.004 & 0.003     \\ 
disc\_ks\_test & 0.062 & 0.065 & 0.099 & 0.089 & 0.107 \\
dis.kstest & 0.003 (536NA) & 0.004 (638NA) & 0.003 (903NA) & 0.001 (996NA) & All NA  \\
kstest.A & 0 & 0 & 0 & 0 & 0   \\
kstest.B & 0 & 0 & 0 & 0 & 0 \\\bottomrule
\end{tabular}
\caption{Type I error rates of KS tests on whether the data comes from ZINB, based on $B=1000$ simulated ZINB datasets with parameters $\phi=0.3$, $r=5$, $p=0.2$ for each sample size $N$}
\label{tab:zinb/zinb}
\end{table}

\begin{table}[hbt!]
\centering
\footnotesize
\begin{tabular}{@{}llllll@{}}
\toprule
    Sample Size $N$ & $30$ & $50$ & $100$ & $200$ & $500$
    \\ \midrule
ks.test & 0.058 & 0 & 0 & 0.001 & 0     \\ 
disc\_ks\_test & 0.062 & 0.049 & 0.067 & 0.07 & 0.084 \\
dis.kstest & 0.788 (176NA) & 0.889 (89NA) & 0.921 (87NA) & 0.832 (168NA) & 0.671 (383NA)    \\
kstest.A & 0 & 0 & 0 & 0.002 & 0.002     \\
kstest.B & 0.002 & 0 & 0 & 0.003 & 0.001      \\ \bottomrule
\end{tabular}
\caption{Type I error rates of KS tests on whether the data comes from ZIBNB, based on $B=1000$ simulated ZIBNB datasets with parameters $\phi=0.3$, $r=3$, $\alpha_1=3$, $\alpha_2=5$ for each sample size $N$}
\label{tab:zibnb/zibnb}
\end{table}

\begin{table}[hbt!]
\centering
\footnotesize
\begin{tabular}{@{}llllll@{}}
\toprule
    Sample Size $N$ & $30$ & $50$ & $100$ & $200$ & $500$
    \\ \midrule
ks.test & 0 & 0 & 0.001 & 0 & 0     \\ 
disc\_ks\_test & 0.054 & 0.06 & 0.06 & 0.058 & 0.066 \\
dis.kstest & 0.039 (24NA) & 0.05 (34NA) & 0.084 (47NA) & 0.645 (64NA) & 0.947 (53NA)    \\
kstest.A & 0.001 & 0 & 0 & 0 & 0     \\
kstest.B & 0.001 & 0 & 0 & 0 & 0      \\ \bottomrule
\end{tabular}
\caption{Type I error rates of KS tests on whether the data comes from ZIBB, based on $B=1000$ simulated ZIBB datasets with parameters $\phi=0.3$, $n=5$, $\alpha_1=8$, $\alpha_2=3$ for each sample size $N$}
\label{tab:zibb/zibb}
\end{table}

\subsection{Comparison study in terms of test power}\label{sec:simulation_study_Algorithm1}

In this section, we continue the simulation studies in Section~\ref{sec:simulation_study_Algorithm1_type_I} and compare the power of the tests based on four different {\tt R} functions. More specifically, given the $B=1000$ datasets simulated from, for example, the ZIP distribution, we run a KS test on whether the data comes from a ZINB distribution, and denote the test as ZIP versus ZINB. If we reject ZINB distribution for $800$ times, then the empirical power of the KS test at ZINB is $800/1000 = 0.80$. Given that the type I error rate does not go beyond the nominal level $0.05$, the bigger the power is, the better the test performs (see, for example, \cite{lehmann2005testing}). We remove function {\tt dis.kstest} of package {\tt iZID} from power analysis since it generates too many {\tt NA}'s for ZINB (see Table~\ref{tab:zinb/zinb}) or has a type I error rate much higher than $0.05$ for ZIBNB (see Table~\ref{tab:zibnb/zibnb}) and ZIBB (see Table~\ref{tab:zibb/zibb}).  

Table~\ref{table-zipvsothers} shows that our functions, {\tt kstest.A} and {\tt kstest.B}, have larger power at ZIP versus ZIBNB. 
All tests fail at ZIP versus ZINB and ZIP versus ZIBB. It often happens for a test of a simpler model versus a more flexible model, especially when the flexible model could approximate the simpler model well (see Section~\ref{sec:discussion} for a discussion on it). More simulation studies show that our functions have the largest powers at ZIBNB versus ZIP, and at ZIBB versus ZIBNB as well (see Tables~\ref{table-zinbvsothers}, \ref{table-zibnbvsothers}, \ref{table-zibbvsothers} in the Supplementary Materials, Section~\ref{sec:more_tables}).
Note that both {\tt ks.test} and {\tt disc\_ks\_test} here rely on the MLEs for ZIBB and ZIBNB provided by our package {\tt AZIAD}. 

Overall our two functions {\tt kstest.A} and {\tt ks.test.B} are most reliable for KS tests involving zero-inflated models. In practice, we recommend {\tt ks.testB} for small sample sizes such as $N=30,50$ (see ZIP versus ZIBNB in Table~\ref{table-zipvsothers}, ZINB versus ZIP in Table~\ref{table-zibnbvsothers}, and ZIBNB versus ZIP in Table~\ref{table-zibnbvsothers}) and {\tt ks.testA} for large sample size such as $N=100,200,500$, which is faster.

\begin{table}[hbt!]
\centering
\footnotesize
\begin{tabular}{@{}lllllll@{}}
\toprule
    Test &  Function & $N=30$ & $N=50$ & $N=100$ & $N=200$ & $N=500$
    \\ \midrule
ZIPvsZINB &ks.test & 0.002 & 0 & 0.001 & 0.001 & 0     \\ 
&disc\_ks\_test & 0.079 & 0.087 & 0.076 & 0.096 & 0.093 \\
& kstest.A & 0 & 0 & 0 & 0 & 0     \\
&kstest.B & 0 & 0 & 0 & 0 & 0      \\ \bottomrule
ZIPvsZIBNB &ks.test & 0.132& 0.441 & 0.844 & 0.978 & 0.984     \\ 
&disc\_ks\_test & 0.115 & 0.096 & 0.084 & 0.099 & 0.101 \\
& kstest.A & 0.749 & 0.892 & 0.954 & 0.973 & 0.992     \\
&kstest.B & 0.73 & 0.984 & 0.954 & 0.973 & 0.992      \\ \bottomrule
ZIPvsZIBB & ks.test & 0.002 & 0.001 & 0.001 & 0.005 & 0.007     \\ 
&disc\_ks\_test & 0.079 & 0.084 & 0.09 & 0.074 & 0.091 \\
&kstest.A & 0 & 0 & 0 & 0.001 & 0     \\
&kstest.B & 0 & 0 & 0 & 0 & 0      \\\bottomrule
\end{tabular}
\caption{Empirical power of KS tests on whether the data comes from ZINB, ZIBNB or ZIBB, based on $B=1000$ simulated ZIP datasets with parameters $\phi=0.3$, $\lambda=10$ for each sample size $N$}
\label{table-zipvsothers}
\end{table}	


\subsection{Real data analysis}\label{sec:realdata}

\begin{example}\label{ex:DedTrivedi}
\textbf{DedTrivedi Data}\quad {\rm
In this example, we use a real data {\tt DebTrivedi} from {\tt R} package {\tt MixAll} to illustrate how to use our package {\tt AZIAD} to identify the most appropriate zero-inflated model. 
The {\tt DebTrivedi} data was obtained from the US National Medial Expenditure Survey \citep{deb1997demand, zeileis2008regression}. It contains $19$ variables from $4,406$ individuals aged $66$ and over. For illustration purpose, we select the variable {\tt ofp}, which is the number of physician office visits. 

In order to analyze the data, we first use KS tests to check each distribution covered by our package. Since the sample size $4,406$ is fairly large, we use our function {\tt kstest.A} instead of {\tt kstest.B}. Out of $19$ distributions under test, we have $7$ models with $p$-value larger than $0.05$, including geometric, NB (with real-valued $r$), NB1 (with integer-valued $r$), ZIBB, ZIBNB, BBH, and BNBH. 


Since there are seven distributions passing the KS test, we further run our likelihood ratio test function {\tt lrt.A} to compare each pair of the candidate distributions. For example, if {\tt d1} represents geometric distribution and {\tt d2} represents NB distribution, we run {\tt lrt.A(d1, d2)}. A $p$-value less than $0.05$ indicates that {\tt d2} is significantly better than {\tt d1}. 
The relevant {\tt R} codes are relegated to the Supplementary Materials (Section~\ref{sec:more_codes}).

Table~\ref{table:lrt.A} shows the results of pairwise comparisons of the seven candidate distributions based on {\tt lrt.A(H0,H1)} with row names indicating $H_0$ and column names indicating $H_1$. A small $p$-value implies that the column distribution is significantly better than the row distribution. 
It should be noted that in general the $p$-values of {\tt lrt.A(d1,d2)} and {\tt lrt.A(d2,d1)} are not equal. 
Based on the $p$-values in Table~\ref{table:lrt.A}, we conclude that ZIBNB and BNBH are significantly better than geometric, NB, NB1, ZIBB and BBH, while there is no significant difference between ZIBNB and BNBH, which confirms our conclusion in Section~\ref{sec:za_vs_zi}.
Note that neither ZIBNB nor BNBH was considered by \cite{deb1997demand} and \cite{zeileis2008regression}. Both distributions were recommended by \cite{aldirawi2019identifying} and \cite{aldirawi2022modeling} for microbiome data analysis. 

We run the same procedure based on our function {\tt lrt.B} as well, which is based on {\tt kstest.B}. 
The results are consistent with {\tt lrt.A}'s. Similarly as in Section~\ref{sec:simulation_study_Algorithm1}, we recommend {\tt lrt.B} for cases with smaller sample sizes and {\tt lrt.A} for cases with a sample size larger than $100$.
}\hfill{$\Box$}
\end{example}

\begin{table}
	\centering
		\begin{tabular}{@{}lllllllll@{}}
			\toprule
		\backslashbox{$H_0$}{$H_1$} & Geometric   & NB   & NB1  & ZIBB  & ZIBNB & BBH & BNBH \\ \bottomrule
			Geometric  & 1 & 0.88 & 0.79 &  0.9 & 0 & 0.81 & 0 \\ \bottomrule
			NB  & 0.07 & 1 & 0.165 &  0.835 & 0 & 0.49 & 0 \\ \bottomrule
			NB1  & 0.84 & 0.82 & 1 &  0.895 & 0 & 0.8 & 0 \\ \bottomrule
			ZIBB  & 0 & 0.04 & 0.015 &  1 & 0 & 0.96 & 0 \\ \bottomrule
			ZIBNB  & 0.66 & 0.725 & 0.69 &  0.81 & 1 & 0.83 & 1 \\ \bottomrule
			BBH  & 0.005 & 0.09 & 0.015 &  1 & 0 & 1 & 0 \\ \bottomrule
			BNBH  & 0.705 & 0.73 & 0.67 &  0.81 & 0.995 & 0.735 & 1 \\ \bottomrule
		\end{tabular}
		\caption{The $p$-values of pairwise comparisons of the seven candidate models based on {\tt lrt.A} for data {\tt DebTrivedi}, a $p$-value less than $0.05$ indicating that the corresponding column model is significantly better than the corresponding row model}
		\label{table:lrt.A}
	\end{table}

\begin{example}\label{ex:omic}
\textbf{Omic Data}\quad
{\rm
In this example, we analyze the {\tt Omic} data from \cite{tipton2018fungi}, which is a list of 229 bacterial and fungal OTUs, for identifying appropriate models. More specifically, for each of the following distributions, Poisson, geometric, negative binomial, beta binomial, beta negative binomial, and their corresponding zero-inflated and hurdle models, we use our {\tt kstest.A} in {\tt AZIAD} to check how many species out of 229 passed the corresponding KS tests. 

Table~\ref{table:kstest.A.omic} summarizes the numbers and percentages of species that do not show significant divergence ($p$-value $> 0.05$). The bigger the number is, the more appropriate the model is for omic data. The relevant {\tt R} codes are displayed below:
\begin{verbatim}
R> dvect = list("poisson", "zip", "ph", "geometric", "zigeom", "geomh", 
      "nb", "zinb", "nbh", "bb", "zibb", "bbh", "bnb", "zibnb", "bnbh")
R> dmatnew = matrix(, nrow=229, ncol=15)
R> set.seed(473415)
R> for(i in 1:229) for(j in 13:15)
  {dmatnew[i,j] = kstest.A(as.numeric(omic[i,]),dist=dvect[j])$pvalue}
R> write.csv(dmatnew, file="kstestBunpolishedfinal.csv")
\end{verbatim}
We conclude that Poisson, geometric, negative binomial (NB), zero-inflated Poisson (ZIP), Poisson hurdle (PH) and geometric hurdle (GeH) are not appropriate for sparse microbial features due to their low percentages (no more than $2\%$),
while beta binomial (BB), beta negative binomial (BNB) and their zero-inflated and hurdle versions are much more popular (at least $64\%$). Among them, BNBH (or ZABNB, see Example~\ref{ex:ZABNB}) is the most appropriate model with percentage $87\%$. Compared with Table~1 in \cite{aldirawi2019identifying} or Table~2 in \cite{aldirawi2022modeling}, our results are more reliable although the patterns are similar.
}\hfill{$\Box$}
\end{example}

\begin{table}
	\centering
		\begin{tabular}{@{}lrr@{}}
			\toprule
		    Distribution & Number & Percentage  \\ \bottomrule
			Poisson  & 0 & 0\% \\ \bottomrule
			Geometric & 2 &  0.8\%  \\\bottomrule
			NB  & 0 & 0\%  \\ \bottomrule
			BB  & 149 & 65\%  \\ \bottomrule
			BNB  & 170 & 74\%  \\ \bottomrule
			ZIP  & 5 & 2\%  \\ \bottomrule
			ZIGe  & 56 & 24\%  \\ \bottomrule
			ZINB  & 57 & 25\%  \\ \bottomrule
			ZIBB  & 148 & 65\%  \\ \bottomrule
			ZIBNB  & 172 & 75\%  \\ \bottomrule
			PH  & 4 & 2\%  \\ \bottomrule
			GeH  & 55  & 24\%  \\ \bottomrule
			NBH  & 56 & 24\%  \\ \bottomrule
			BBH  & 181 & 79\%  \\ \bottomrule
			BNBH  & 200 & 87\%  \\ \bottomrule
			
		\end{tabular}
		\caption{Number and Percentage of species out of 229 that passed {\tt kstest.A} with  $p$-value $>0.05$}
		\label{table:kstest.A.omic}
	\end{table}

\section{Discussion}\label{sec:discussion}

A major goal targeted in this paper is to identify the underlying distribution $F_0$ given a random sample $\{Y_1, \ldots, Y_n\}$ from it. Given the data $\{Y_1, \ldots, Y_n\}$, ideally our tests could accomplish two tasks, (i) do not reject $F_0$ itself; (ii) reject any $F_1$ which is not $F_0$. 

According to the Glivenko-Cantelli theorem (see, for example, Theorem~19.1 in \cite{vaart2000asymptotic}), the empirical distribution function $F_n$ converges to $F_0$ uniformly and then Task~(i) can be achieved by proper KS tests up to a type I error, which is confirmed by our simulation studies in Section~\ref{sec:simulation_study_Algorithm1_type_I}.

Task~(ii) is much more complicated. First of all, in Section~\ref{sec:za_vs_zi}, we show the equivalence of a zero-inflated model and its corresponding hurdle model given that $\phi_{\rm ZA} \geq p_0(\boldsymbol\theta)$. Therefore, one could not make a distinction between a ZI model and its corresponding ZA or hurdle model when zero-inflation exists. We have such a real data example in Section~\ref{sec:realdata}. 

Secondly, as shown by simulation studies in Section~\ref{sec:simulation_study_Algorithm1}, one may not be able to reject $F_1$ if $F_1$ is a more flexible model than $F_0$, especially when $F_1$ with fitted parameters could approximate $F_0$ well. In this section, we use ZIP versus ZINB as an example to illustrate why one may not be able to reject $F_1$ given that $F_0$ is the true model.

It is known that Poisson($\lambda$) can be approximated by NB($r,p$) with a large $r$ and a $p$ close to $1$ such that $\lambda = r(1-p)$ (see, for example, \cite{teerapabolarn2014}).
To illustrate how well ZINB could approximate a ZIP distribution, we simulate random samples $\{Y_1, \ldots, Y_N\}$ from ZIP($\phi=0.3, \lambda = 10$) with various sample sizes $N$, then fit a ZINB model. In Table~\ref{table:ZIP-ZINB}, we list the maximum difference between the cumulative distribution function (CDF) $F_{\rm ZIP}(y)$ of ZIP and the CDF $F_{\rm ZINB}(y)$ of the fitted ZINB. As the sample size $N$ increases, the maximum distance $\sup_y |F_{\rm ZIP}(y) - F_{\rm ZINB}(y)|$ decreases, which indicates ZINB approximates ZIP better and better.

	\begin{table}
	\centering
		\begin{tabular}{@{}llllllll@{}}
			\toprule
			$N$ & $30$   & $50$   & $100$  & $200$  & $500$ & $1000$ & $5000$  \\ \bottomrule
			$\sup_y |F_{\rm ZIP} (y) - F_{\rm ZINB}(y)|$  & 0.166 & 0.12 & 0.08 &  0.03 & 0.024 & 0.04 & 0.0158 \\ \bottomrule
		\end{tabular}
		\caption{Maximum difference between the CDF of ZIP and the CDF of fitted ZINB based on random samples from ZIP($\phi = 0.3, \lambda = 10$) with various sample sizes $N$}
		\label{table:ZIP-ZINB}
	\end{table}

Another example is the KS test of ZIBNB versus ZIBB, whose empirical power shows a decreasing pattern as the sample size $N$ increases (see Table~\ref{table-zibnbvsothers} in the Supplementary Materials). We perform a similar numerical study here for ZIBNB versus ZIBB as well. More specifically, we simulate random samples $\{Y_1, \ldots, Y_N\}$ from ZIBNB($\phi=0.3, r=15, \alpha_1=19, \alpha_2=10$) with various sample sizes $N$, then fit a ZIBB model. Table~\ref{table:ZIBNB-ZIBB} shows a decreasing pattern of the maximum difference between the two CDFs as the sample size $N$ increases, which indicates ZIBB can approximate ZIBNB better and better.

	\begin{table}
	\centering
		\begin{tabular}{@{}llllllll@{}}
			\toprule
			$N$ & $30$   & $50$   & $100$  & $200$  & $500$ & $1000$ & $5000$  \\ \bottomrule
			$\sup_y |F_{\rm ZIBNB}(y)-F_{\rm ZIBB}(y)|$  & 0.166 & 0.1 & 0.1 & 0.11 & 0.04 & 0.024 & 0.014 \\ \bottomrule
		\end{tabular}
		\caption{Maximum difference between the CDF of ZIBNB and the CDF of fitted ZIBB based on random samples from ZIBNB($\phi = 0.3, r=15, \alpha_1=19, \alpha_2=10$) with various sample sizes $N$}
		\label{table:ZIBNB-ZIBB}
	\end{table}

\section*{Acknowledgments}

This work was supported in part by the U.S.~NSF grant DMS-1924859, and CSUSB research fellowship.

\clearpage
\setcounter{page}{1}
\def\thepage{S\arabic{page}}

\fontsize{10.95}{14pt plus.8pt minus .6pt}\selectfont
\vspace{0.8pc}
\begin{center}
{\large\bf An R Package AZIAD for Analyzing Zero-Inflated and Zero-Altered Data}
\end{center}
\vspace{.25cm}
\centerline{Niloufar Dousti Mousavi$^{a}$, Hani Aldirawi$^{b}$ and Jie Yang$^{a}$}
\vspace{.4cm}
\centerline{\it  $^a$University of Illinois at Chicago~and $^b$California State University, San Bernardino}
\vspace{.55cm}
 \centerline{\bf Supplementary Materials}
\vspace{.55cm}
\fontsize{9}{11.5pt plus.8pt minus .6pt}\selectfont
\par

\renewcommand{\thesection}{S}
\setcounter{equation}{0}
\setcounter{subsection}{0}
\renewcommand{\theequation}{S.\arabic{equation}}
\setcounter{figure}{0}
\renewcommand{\thefigure}{S.\arabic{figure}}
\setcounter{table}{0}
\renewcommand{\thetable}{S.\arabic{table}}

\subsection{Proofs}\label{sec:proofs}

\medskip\noindent
{\bf Proof of Theorem~\ref{thm:inverse_fisher_ZI}:}
According to  formulae~14.13(b) in \cite{seber2008}, if $\beta=a-{\mathbf b}^T {\mathbf D}^{-1} {\mathbf b} \neq 0$, then
\begin{equation}\label{eq:block_matrix_inverse}
\begin{bmatrix}
a & {\mathbf b}^T \\
{\mathbf b} & {\mathbf D}
\end{bmatrix}^{-1} = 
\begin{bmatrix}
\frac{1}{\beta} & -\frac{1}{\beta}{\mathbf b}^T {\mathbf D}^{-1}\\
-\frac{1}{\beta} {\mathbf D}^{-1} {\mathbf b}  & {\mathbf D}^{-1}+\frac{1}{\beta} {\mathbf D}^{-1} {\mathbf b} {\mathbf b}^T {\mathbf D}^{-1}
\end{bmatrix} =
\begin{bmatrix}
\frac{1}{\beta} & -\frac{1}{\beta}\boldsymbol\gamma^T \\
-\frac{1}{\beta}\boldsymbol\gamma  & {\mathbf D}^{-1}+\frac{1}{\beta}\boldsymbol\gamma \boldsymbol\gamma^T
\end{bmatrix}
\end{equation}
where $a\in \mathbb{R}$, ${\mathbf b}$ is a vector, ${\mathbf D}$ is a symmetric nonsingular matrix, $\boldsymbol\gamma = {\mathbf D}^{-1} {\mathbf b}$ is a vector.
For our case ${\mathbf F}_{\rm ZI}$ as in \eqref{eq:fisher_ZI}, $a = \frac{1-p_0(\boldsymbol\theta)}{[\phi + (1-\phi)p_0(\boldsymbol\theta)] (1-\phi)}$, ${\mathbf b} = \frac{p_0(\boldsymbol\theta)}{\phi + (1-\phi) p_0(\boldsymbol\theta)} \cdot \frac{\partial \log p_0(\boldsymbol\theta)}{\partial \boldsymbol\theta}$, ${\mathbf D} = {\mathbf F}_{\rm ZI\boldsymbol\theta} = (1-\phi) ({\mathbf F}_{\boldsymbol\theta} - \frac{\phi p_0(\boldsymbol\theta)}{\phi + (1-\phi) p_0(\boldsymbol\theta)} \cdot \frac{\partial \log p_0(\boldsymbol\theta)}{\partial \boldsymbol\theta}\cdot \frac{\partial \log p_0(\boldsymbol\theta)}{\partial \boldsymbol\theta^T})$. Since $|{\mathbf F}_{\rm ZI}| \neq 0$, then ${\mathbf F}_{\rm ZI}$ is positive definite and so is ${\mathbf D}$. On the other hand, $|{\mathbf F}_{\rm ZI}| = |{\mathbf D}|\cdot (a - {\mathbf b}^T {\mathbf D}^{-1} {\mathbf b}) = |{\mathbf D}|\cdot \beta$ (see, for example, formula~14.21(b) in \cite{seber2008}), then $|{\mathbf F}_{\rm ZI}|\neq 0$ implies $\beta \neq 0$ in our case.

To simplify the notations, we write in this proof only $p = p_{0}(\boldsymbol\theta)$, $d = d_{\boldsymbol\theta} = \phi + (1-\phi)p_0(\boldsymbol\theta)$, and ${\mathbf F} =  {\mathbf F}_{\boldsymbol\theta}$.  
We first derive the formula of ${\mathbf D}^{-1}$. 
By the Sherman-Morrison-Woodbury formula (see for example, 15.3(a)(iii) in \cite{seber2008}), $(A+UV)^{-1} = A^{-1} - A^{-1}U (I + VA^{-1}U)^{-1}VA^{-1}$ if $|A| \neq 0$ and $|I + VA^{-1}U|\neq 0$. In our case, $A={\mathbf F}$ satisfies $|A| = |{\mathbf F}| = |{\mathbf F}_{\boldsymbol\theta}|\neq 0$, $U = -\frac{\phi p}{d} \frac{\partial\log p}{\partial\boldsymbol\theta}$, $V= \frac{\partial\log p}{\partial\boldsymbol\theta^T}$. On the other hand, $|{\mathbf D}|\neq 0$ implies $\phi < 1$ and $|{\mathbf F} - \frac{\phi p}{d} \cdot \frac{\partial \log p}{\partial \boldsymbol\theta} \cdot \frac{\partial \log p}{\partial \boldsymbol\theta^T}| \neq 0$. Since $|{\mathbf F}| \neq 0$, then ${\mathbf F}$ is positive definite and ${\mathbf F}^{-1/2}$ exists. According to formulae~4.33 in \cite{seber2008}, $|I - {\mathbf a}{\mathbf a}^T| = 1-{\mathbf a}^T{\mathbf a}$ for any vector ${\mathbf a}$. Then
\begin{eqnarray*}
0 &\neq & |{\mathbf F}^{-1/2}|\cdot \left|{\mathbf F} - \frac{\phi p}{d} \cdot \frac{\partial \log p}{\partial \boldsymbol\theta} \cdot \frac{\partial \log p}{\partial \boldsymbol\theta^T}\right| \cdot |{\mathbf F}^{-1/2}|\\
&=& \left|I - \left(\sqrt{\frac{\phi p}{d}} {\mathbf F}^{-1/2} \frac{\partial \log p}{\partial \boldsymbol\theta}\right)\cdot \left(\sqrt{\frac{\phi p}{d}} {\mathbf F}^{-1/2} \frac{\partial \log p}{\partial \boldsymbol\theta}\right)^T\right|\\
&=& 1- \left(\sqrt{\frac{\phi p}{d}} {\mathbf F}^{-1/2} \frac{\partial \log p}{\partial \boldsymbol\theta}\right)^T \cdot \left(\sqrt{\frac{\phi p}{d}} {\mathbf F}^{-1/2} \frac{\partial \log p}{\partial \boldsymbol\theta}\right)\\
&=& 1- \frac{\phi p}{d} \frac{\partial \log p}{\partial \boldsymbol\theta^T} {\mathbf F}^{-1} \frac{\partial \log p}{\partial \boldsymbol\theta}\\
&=& |I + VA^{-1}U|
\end{eqnarray*}
That is, $|I + VA^{-1}U|\neq 0$ in our case. By applying the Sherman-Morrison-Woodbury formula, 
\[
{\mathbf D}^{-1} =  \frac{1}{1-\phi}\left({\mathbf F} - \frac{\phi p}{d} \cdot \frac{\partial \log p}{\partial \theta} \cdot \frac{\partial \log p}{\partial \theta^T}\right)^{-1}
= \frac{1}{1-\phi}\left({\mathbf F}^{-1} + \frac{\phi p}{d\delta} {\mathbf c} {\mathbf c}^T\right)
\]
where $\delta = \delta_{\boldsymbol\theta} = 1- \frac{\phi p}{d} \frac{\partial \log p}{\partial \boldsymbol\theta^T} {\mathbf F}^{-1} \frac{\partial \log p}{\partial \boldsymbol\theta} \neq 0$ and ${\mathbf c} = {\mathbf F}^{-1} \frac{\partial \log p}{\partial \theta}$. It can be verified that $\boldsymbol\gamma = {\mathbf D}^{-1} {\mathbf b} = \frac{p}{d\delta (1-\phi)} \cdot {\mathbf c}$ and $\beta = a - {\mathbf b}^T {\mathbf D}^{-1} {\mathbf b} = \frac{1}{\phi(1-\phi)}\cdot \frac{d\delta - p}{d\delta} \neq 0$. Then the final expression of ${\mathbf F}_{\rm ZI}^{-1}$ can be obtained from \eqref{eq:block_matrix_inverse}.
\hfill{$\Box$}

\medskip\noindent
{\bf Proof of Theorem~\ref{thm:za_vs_zi}:}
Given $f_{\rm ZI}(y\mid \phi_{\rm ZI}, \boldsymbol\theta)$ and $\phi_{\rm ZA} = \phi_{\rm ZI} + (1-\phi_{\rm ZI}) p_0(\boldsymbol\theta)$, we have $1-\phi_{\rm ZA} = (1-\phi_{\rm ZI})[1-p_0(\boldsymbol\theta)]$ and $f_{\rm ZA}(y\mid \phi_{\rm ZA}, \boldsymbol\theta) = \phi_{\rm ZA} {\mathbf 1}_{\{y=0\}} + (1-\phi_{\rm ZA}) \frac{f_{\boldsymbol\theta}(y)}{1-p_0(\boldsymbol\theta)} {\mathbf 1}_{\{y \neq 0\}} = [\phi_{\rm ZI} + (1-\phi_{\rm ZI}) p_0(\boldsymbol\theta)] {\mathbf 1}_{\{y = 0\}} + (1-\phi_{\rm ZI}) f_{\boldsymbol\theta}(y) {\mathbf 1}_{\{y\neq 0\}} =  f_{\rm ZI}(y\mid \phi_{\rm ZI}, \boldsymbol\theta)$.

Given $f_{\rm ZA}(y\mid \phi_{\rm ZA}, \boldsymbol\theta)$ and $\phi_{\rm ZI} = [\phi_{\rm ZA} - p_0(\boldsymbol\theta)]/[1- p_0(\boldsymbol\theta)]$, if $\phi_{\rm ZA} \geq p_0(\boldsymbol\theta)$, then $\phi_{\rm ZI} \in [0,1]$, $\phi_{\rm ZI} + (1-\phi_{\rm ZI}) p_0(\boldsymbol\theta) = \phi_{\rm ZA}$, $1-\phi_{\rm ZI} = \frac{1-\phi_{\rm ZA}}{1-p_0(\boldsymbol\theta)}$, and $f_{\rm ZI}(y\mid \phi_{\rm ZI}, \boldsymbol\theta) = \phi_{\rm ZA} {\mathbf 1}_{\{y=0\}} + \frac{1-\phi_{\rm ZA}}{1-p_0(\boldsymbol\theta)} f_{\boldsymbol\theta}(y) {\mathbf 1}_{\{y\neq 0\}} = \phi_{\rm ZA} {\mathbf 1}_{\{y=0\}} + (1-\phi_{\rm ZA}) f_{\rm tr}(y\mid \boldsymbol\theta) {\mathbf 1}_{\{y\neq 0\}} = f_{\rm ZA}(y\mid \phi_{\rm ZA}, \boldsymbol\theta)$.
\hfill{$\Box$}

\subsection{More examples}\label{sec:more_examples}

\begin{example}\label{ex:ZIBB}{\bf Zero-inflated beta-binomial model (ZIBB)} {\rm Same as in Example~\ref{ex:ZABB}, the pmf of the baseline distribution is $f_{\boldsymbol\theta} (y) = {n\choose y} \frac{{\rm Beta}(y + \alpha, n - y + \beta)}{{\rm Beta}(\alpha,\beta)}$ with parameters $\boldsymbol\theta = (n, \alpha, \beta)$ and $y=0, 1, \ldots, n$. 
According to Theorem~5 in \cite{aldirawi2022modeling}, the Fisher information matrix of the ZIBB distribution is 
\[
{\mathbf F}_{\rm ZIBB} =  \begin{bmatrix}
C_{11} & C_{12} & C_{13} & C_{14}\\
C_{12} & & & \\
C_{13} & & \begin{matrix}
{\mathbf F}_{{\rm ZIBB}\boldsymbol\theta}
\end{matrix}\\
C_{14} & & &
\end{bmatrix}
\]
where
\begin{eqnarray*}
C_{11}&=&\frac{\Gamma(n+\alpha+\beta)\Gamma(\beta)-\Gamma(n+\beta)\Gamma(\alpha+\beta)}{[\phi\Gamma(n+\alpha+\beta)\Gamma(\beta)+
(1-\phi)\Gamma(n+\beta)\Gamma(\alpha+\beta)](1-\phi)}\\
C_{12}&=& \frac{\Gamma(n+\beta)\Gamma(\alpha+\beta)[\Psi(n+\beta) - \Psi(n+\alpha+\beta)]}{\phi\Gamma(n+\alpha+\beta)\Gamma(\beta)+
(1-\phi)\Gamma(n+\beta)\Gamma(\alpha+\beta)} \\
C_{13}&=& \frac{\Gamma(n+\beta)\Gamma(\alpha+\beta)[\Psi(\alpha+\beta) - \Psi(n+\alpha+\beta)]}{\phi\Gamma(n+\alpha+\beta)\Gamma(\beta)+
(1-\phi)\Gamma(n+\beta)\Gamma(\alpha+\beta)}\\
C_{14} &=& \frac{\Gamma(n+\beta)\Gamma(\alpha+\beta)[\Psi(n+\beta) + \Psi(\alpha+\beta) - \Psi(n+\alpha+\beta) - \Psi(\beta)]}{\phi\Gamma(n+\alpha+\beta)\Gamma(\beta)+
(1-\phi)\Gamma(n+\beta)\Gamma(\alpha+\beta)}\\ 
{\mathbf F}_{{\rm ZIBB}\boldsymbol\theta} &=& -(1-\phi) \left({\mathbf A} +\frac{\phi\Gamma(n+\beta)\Gamma(\alpha+\beta)}{\phi\Gamma(n+\alpha+\beta)\Gamma(\beta)+ (1-\phi)\Gamma(n+\beta)\Gamma(\alpha+\beta)}\cdot {\mathbf B}\right)
\end{eqnarray*}
Note that 
\[
{\mathbf A} = \begin{bmatrix}
A_{11} & A_{12} & A_{13}\\
A_{12} & A_{22} & A_{23}\\
A_{13} & A_{23} & A_{33}
\end{bmatrix},\>\>\>
{\mathbf B} = 
\begin{bmatrix}
B_{11} & B_{12} & B_{13}\\
B_{21} & B_{22} & B_{23}\\
B_{31} & B_{32} & B_{33}
\end{bmatrix}
\]
here are the same as in Example~\ref{ex:ZABB}.
}\hfill{$\Box$}
\end{example}

\begin{example}\label{ex:ZIBNB} {\bf Zero-inflated beta negative binomial model (ZIBNB)} {\rm Same as in Example~\ref{ex:ZABNB}, the pmf of the baseline distribution  is given by $f_{\boldsymbol\theta}(y) = {r+y-1\choose y} \frac{{\rm Beta}(r + \alpha,  y + \beta)}{{\rm Beta}(\alpha,\beta)}$  with parameters $\boldsymbol\theta = (r, \alpha, \beta)$, $y \in \{0, 1, 2, \ldots\}$. 
According to Theorem~5 in \cite{aldirawi2022modeling}, the Fisher information matrix of the ZIBNB distribution is 
\[
{\mathbf F}_{\rm ZIBNB} = \begin{bmatrix}
C_{11} & C_{12} & C_{13} & C_{14}\\
C_{12} & & & \\
C_{13} & & \begin{matrix}
{\mathbf F}_{{\rm ZIBNB}\boldsymbol\theta}
\end{matrix}\\
C_{14} & & &
\end{bmatrix}
\]
where
\begin{eqnarray*}
C_{11}&=&\frac{\Gamma(r+\alpha+\beta)\Gamma(\alpha)-\Gamma(r+\alpha)\Gamma(\alpha+\beta)}{[\phi\Gamma(r+\alpha+\beta)\Gamma(\alpha)+
(1-\phi)\Gamma(r+\alpha)\Gamma(\alpha+\beta)](1-\phi)}\\
C_{12}&=& \frac{\Gamma(r+\alpha)\Gamma(\alpha+\beta)[\Psi(r+\alpha) - \Psi(r+\alpha+\beta)]}{\phi\Gamma(r+\alpha+\beta)\Gamma(\alpha)+
(1-\phi)\Gamma(r+\alpha)\Gamma(\alpha+\beta)} \\
C_{13}&=& \frac{\Gamma(r+\alpha)\Gamma(\alpha+\beta)[\Psi(r+\alpha) + \Psi(\alpha+\beta) - \Psi(r+\alpha+\beta) - \Psi(\alpha)]}{\phi\Gamma(r+\alpha+\beta)\Gamma(\alpha)+
(1-\phi)\Gamma(r+\alpha)\Gamma(\alpha+\beta)}\\
C_{14} &=& \frac{\Gamma(r+\alpha)\Gamma(\alpha+\beta)[\Psi(\alpha+\beta) - \Psi(r+\alpha+\beta)]}{\phi\Gamma(r+\alpha+\beta)\Gamma(\alpha)+
(1-\phi)\Gamma(r+\alpha)\Gamma(\alpha+\beta)}\\
{\mathbf F}_{{\rm ZIBNB}\boldsymbol\theta} &=& -(1-\phi) \left({\mathbf A} + \frac{\phi\Gamma(r+\alpha)\Gamma(\alpha+\beta)}{\phi\Gamma(r+\alpha+\beta)\Gamma(\alpha)+
(1-\phi)\Gamma(r+\alpha)\Gamma(\alpha+\beta)} \cdot {\mathbf B}\right)
\end{eqnarray*}
Note that 
\[
{\mathbf A} = \begin{bmatrix}
A_{11} & A_{12} & A_{13}\\
A_{12} & A_{22} & A_{23}\\
A_{13} & A_{23} & A_{33}
\end{bmatrix},\>\>\>
{\mathbf B} = \begin{bmatrix}
B_{11} & B_{12} & B_{13}\\
B_{12} & B_{22} & B_{23}\\
B_{13} & B_{23} & B_{33}
\end{bmatrix}
\]
are the same as in Example~\ref{ex:ZABNB}.
}\hfill{$\Box$}
\end{example}

\subsection{More R codes}\label{sec:more_codes}

For readers' reference, we provide the {\tt R} code for generating Table~\ref{tab:zinb/zinb}.

\begin{verbatim}
R> set.seed(007)
R> NoIterations <- 1000
R> listnb1 <- numeric(NoIterations)
R> listnb2 <- numeric(NoIterations)
R> listnb3 <- numeric(NoIterations)
R> listnb4 <- numeric(NoIterations)
R> listnb5 <- numeric(NoIterations)
R> for(i in 1:NoIterations){
R> sampled <-  sample.zi(N=500,phi=0.3,dist = 'nb',r=5,p=.2)
R> listnb1[i]<-dist.ksnew(sampled,nsim=100,bootstrap = TRUE,dist= 'zinb',
                          lowerbound = 1e-10, upperbound = 100000)$pvalue
R> listnb2[i]<-dist.ksmle(sampled,nsim=100,bootstrap = TRUE,dist= 'zinb',
                          lowerbound = 1e-10, upperbound = 100000)$pvalue
R> listnb3[i]<-try(dis.kstest(sampled,nsim=100,bootstrap = TRUE,distri='zinb',
                          lowerbound = 1e-10, upperbound = 100000)$pvalue }
R> for(i in 1:NoIterations) {             
R>   sampled <-sample.zi(N=30,phi=0.3,dist = 'nb',r=5,p=.2)
R>   a=zih.mle(sampled,r=10,p=.3,type="zi",dist = "nb1.zihmle")
R>   sampled1<- sample.zi(N=30,r=a[1],p=a[2],phi=a[3],dist="nb")
R>   listnb4[i]<-ks.test(sampled,sampled1)$p.value
R>   x.unique=sort(unique(sampled))
R>   probs_ori = a[3] + (1 - a[3]) * stats::pnbinom(x.unique, 
                                    size = a[1], prob = a[2])
R>   step_ori = stats::stepfun(x.unique, c(0, probs_ori))
R>   listnb5[i]<-disc_ks_test(x=sampled,y=step_ori,exact = TRUE)$p.value }
\end{verbatim}

\medskip\noindent
Below we provide the {\tt R} code for analyzing the {\tt DebTrivedi} data in Section~\ref{sec:realdata}.

\begin{verbatim}
R> set.seed(1008)
R> kstest.A(ofp,nsim=200,bootstrap=TRUE,dist="poisson")$pvalue    #0
R> kstest.A(ofp,nsim=200,bootstrap=TRUE,dist="geometric")$pvalue  #0.06
R> kstest.A(ofp,nsim=200,bootstrap=TRUE,dist="nb")$pvalue         #0.08
R> kstest.A(ofp,nsim=200,bootstrap=TRUE,dist="nb1")$pvalue        #0.105
R> kstest.A(ofp,nsim=200,bootstrap=TRUE,dist="bb")$pvalue         #0.02
R> kstest.A(ofp,nsim=200,bootstrap=TRUE,dist="bb1")$pvalue        #0.04
R> kstest.A(ofp,nsim=200,bootstrap=TRUE,dist="bnb")$pvalue        #0.015
R> kstest.A(ofp,nsim=200,bootstrap=TRUE,dist="bnb")$pvalue        #0.025
R> kstest.A(ofp,nsim=200,bootstrap=TRUE,dist="bnb1")$pvalue       #0.01
R> kstest.A(ofp,nsim=200,bootstrap=TRUE,dist="zip")$pvalue        #0
R> kstest.A(ofp,nsim=200,bootstrap=TRUE,dist="zigeom")$pvalue     #0.015
R> kstest.A(ofp,nsim=200,bootstrap=TRUE,dist="zinb")$pvalue       #0.005
R> kstest.A(ofp,nsim=200,bootstrap=TRUE,dist="zibb")$pvalue       #0.085
R> kstest.A(ofp,nsim=200,bootstrap=TRUE,dist="zibnb")$pvalue      #0.85
R> kstest.A(ofp,nsim=200,bootstrap=TRUE,dist="ph")$pvalue         #0
R> kstest.A(ofp,nsim=200,bootstrap=TRUE,dist="geomh")$pvalue      #0.005
R> kstest.A(ofp,nsim=200,bootstrap=TRUE,dist="nbh")$pvalue        #0.005
R> kstest.A(ofp,nsim=200,bootstrap=TRUE,dist="bbh")$pvalue        #0.08
R> kstest.A(ofp,nsim=200,bootstrap=TRUE,dist="bnbh")$pvalue       #0.825
\end{verbatim}

\begin{verbatim}
#Best Model By using LRT
R> d1=kstest.A(ofp,nsim=200,bootstrap=TRUE,dist="geometric")       
R> d2=kstest.A(ofp,nsim=200,bootstrap=TRUE,dist="nb")              
R> d3=kstest.A(ofp,nsim=200,bootstrap=TRUE,dist="nb1")             
R> d4=kstest.A(ofp,nsim=200,bootstrap=TRUE,dist="zibb")            
R> d5=kstest.A(ofp,nsim=200,bootstrap=TRUE,dist="zibnb")           
R> d6=kstest.A(ofp,nsim=200,bootstrap=TRUE,dist="bbh")             
R> d7=kstest.A(ofp,nsim=200,bootstrap=TRUE,dist="bnbh")        
R> pmat = matrix(, nrow=7, ncol=7)
R> dvec = list(d1,d2,d3,d4,d5,d6,d7)
    for(i in 1:7) for(j in 1:7) pmat[i,j] = lrt.A(dvec[[i]],dvec[[j]]);
\end{verbatim}

\subsection{More tables}\label{sec:more_tables}

In this section, we provide more tables relevant to Section~\ref{sec:simulation_study_Algorithm1}.

Table~\ref{table-zinbvsothers} shows that our functions, {\tt kstest.A} and {\tt kstest.B}, have comparable power at ZINB versus ZIP. All tests fail at ZINB versus ZIBNB. All tests lose power gradually at ZINB versus ZIBB,  as sample size $N$ increases. 
Table~\ref{table-zibnbvsothers} shows that our functions have the largest powers at ZIBNB versus ZIP. For ZIBNB versus ZINB, our functions have larger power than {\tt ks.test} but less power than {\tt disc\_ks\_test}, which tends to reject more for ZINB (see Table~\ref{tab:zinb/zinb}). All tests lose power quickly at ZIBNB versus ZIBB, while {\tt disc\_ks\_test} shows a slower decreasing rate. 
Table~\ref{table-zibbvsothers} shows that our functions have significantly larger power at ZIBB versus ZIBNB.
Function {\tt disc\_ks\_test} has greater power at ZIBB versus ZINB and ZIBB versus ZIP when $N \leq 100$. 

\begin{table}[hbt!]
\centering
\footnotesize
\begin{tabular}{@{}lllllll@{}}
\toprule
    Test &  Function & $N=30$ & $N=50$ & $N=100$ & $N=200$ & $N=500$
    \\ \midrule
ZINBvsZIP & ks.test & 0.05 & 0.111 & 0.325 & 0.801 & 1     \\ 
&disc\_ks\_test & 0.316 & 0.54 & 0.89 & 1 & 1 \\
&kstest.A & 0.2 & 0.501 & 0.942 & 1 & 1     \\
&kstest.B & 0.199 & 0.512 & 0.942 & 1 & 1 \\\bottomrule
ZINBvsZIBNB&ks.test & 0 & 0 & 0 & 0 & 0     \\ 
&disc\_ks\_test & 0.054 & 0.045 & 0.045 & 0.041 & 0.056 \\
&kstest.A & 0 & 0 & 0 & 0 & 0     \\
&kstest.B & 0 & 0 & 0 & 0 & 0      \\ \bottomrule
ZINBvsZIBB&ks.test & 0.297 & 0.529 & 0.6 & 0.567 & 0.429     \\ 
&disc\_ks\_test & 0.634 & 0.663 & 0.646 & 0.57 & 0.457 \\ 
&kstest.A & 0.632 & 0.641 & 0.59 & 0.535 & 0.368     \\
&kstest.B & 0.632 & 0.642 & 0.59 & 0.535 & 0.368      \\ \bottomrule
\end{tabular}
\caption{Empirical power of KS tests on whether the data comes from ZIP, ZIBNB or ZIBB, based on $B=1000$ simulated ZINB datasets with parameters $\phi=0.3$, $r=3$, $\alpha_1=3$, $\alpha_2=5$ for each sample size $N$}
\label{table-zinbvsothers}
\end{table}	

\begin{table}[hbt!]
\centering
\footnotesize
\begin{tabular}{@{}lllllll@{}}
\toprule
    Test &  Function & $N=30$ & $N=50$ & $N=100$ & $N=200$ & $N=500$
    \\ \midrule
ZIBNBvsZIP&ks.test & 0.183 & 0.413 & 0.751 & 0.975 & 1     \\ 
&disc\_ks\_test & 0.552 & 0.764 & 0.981 & 1 & 1 \\
&kstest.A & 0.569 & 0.866 & 0.996 & 1 & 1     \\
&kstest.B & 0.573 & 0.871 & 0.996 & 1 & 1      \\\bottomrule
ZIBNBvsZINB&ks.test & 0.006 & 0.011 & 0.022 & 0.021 & 0.099     \\ 
&disc\_ks\_test & 0.106 & 0.14 & 0.145 & 0.253 & 0.437 \\
&kstest.A & 0.019 & 0.016 & 0.043 & 0.097 & 0.248     \\
&kstest.B & 0.018 & 0.019 & 0.046 & 0.102 & 0.307      \\ \bottomrule
ZIBNBvsZIBB&ks.test & 0.422 & 0.266 & 0.167 & 0.069 & 0.046     \\ 
&disc\_ks\_test & 0.473 & 0.337 & 0.251 & 0.209 & 0.285 \\
&kstest.A & 0.413 & 0.282 & 0.144 & 0.084 & 0.056     \\
&kstest.B & 0.413 & 0.282 & 0.144 & 0.084 & 0.056       \\
 \bottomrule
\end{tabular}
\caption{Empirical power of KS tests on whether the data comes from ZIP, ZINB or ZIBB, based on $B=1000$ simulated ZIBNB datasets with parameters $\phi=0.3$, $r=3$, $\alpha_1=3$, $\alpha_2=5$ for each sample size $N$}
\label{table-zibnbvsothers}
\end{table}	

\begin{table}[hbt!]
\centering
\footnotesize
\begin{tabular}{@{}lllllll@{}}
\toprule
    Test &  Function & $N=30$ & $N=50$ & $N=100$ & $N=200$ & $N=500$
    \\ \midrule
ZIBBvsZIP&ks.test & 0.01 & 0.017 & 0.038 & 0.235 & 0.983     \\ 
&disc\_ks\_test & 0.241 & 0.299 & 0.714 & 0.963 & 1 \\
&kstest.A & 0.001 & 0.037 & 0.459 & 0.986 & 1     \\
&kstest.B & 0.01 & 0.05 & 0.455 & 0.992 & 1      \\\bottomrule
ZIBBvsZINB&ks.test & 0.008 & 0.016 & 0.052 & 0.246 & 0.989     \\ 
&disc\_ks\_test & 0.237 & 0.316 & 0.688 & 0.965 & 1 \\
&kstest.A & 0.015 & 0.058 & 0.494 & 0.99 & 1     \\
&kstest.B & 0.015 & 0.063 & 0.494 & 0.99 & 1      \\ \bottomrule
ZIBBvsZIBNB&ks.test & 0.011 & 0.039 & 0.144 & 0.548 & 1     \\ 
&disc\_ks\_test & 0.86 & 0.075 & 0.085 & 0.126 & 0.262 \\
&kstest.A & 0.275 & 0.75 & 0.999 & 1 & 1     \\
&kstest.B & 0.265 & 0.739 & 0.998 & 1 & 1      \\\bottomrule
\end{tabular}
\normalsize
\caption{Empirical power of KS tests on whether the data comes from ZIP, ZINB or ZIBNB, based on $B=1000$ simulated ZIBB datasets with parameters $\phi=0.3$, $n=5$, $\alpha_1=8$, $\alpha_2=3$ for each sample size $N$}
\label{table-zibbvsothers}
\end{table}

\end{document}